\documentclass[apjl]{emulateapj}

\usepackage{graphicx}
\usepackage{t1enc}
\usepackage[varg]{txfonts}
\usepackage{epsfig}
\usepackage{rotating}
\usepackage{natbib}
\usepackage{color}

\newcommand{\Nmm} 	{$N_\mathrm{mm}$} 
\newcommand{\Surfd} 	{$\Sigma_\mathrm{0.05\,pc}$}
\newcommand{\nradius} 	{$n_\mathrm{0.05\,pc}$}
\newcommand{\Lbol}		{$L_\mathrm{bol}$}
\newcommand{\betar}		{$\beta_\mathrm{rot}$}

\newcommand{\mum}   	{$\mu$m}
\newcommand{\kms}   {km~s$^{-1}$}

\newcommand{\cmt}   {cm$^{-3}$}
\newcommand{\jpb}   {$\rm Jy~beam^{-1}$}    
\newcommand{\lo}    {$L_{\sun}$}
\newcommand{\mo}    {$M_{\sun}$}
\newcommand{\nh}    {NH$_3$}

\newcommand{\et}    {et al.}
\newcommand{\eg}    {e.\,g.,}
\newcommand{\ie}    {i.\,e.,}

\newcommand{\supa}  {$^\mathrm{a}$}
\newcommand{\supb}  {$^\mathrm{b}$}
\newcommand{\supc}  {$^\mathrm{c}$}
\newcommand{\supd}  {$^\mathrm{d}$}
\newcommand{\supe}  {$^\mathrm{e}$}

\definecolor{RED}{rgb}{1.0,0.0,0.0}

\shorttitle{Fragmentation of massive cores down to $\sim1000$~AU}
\shortauthors{Palau et al.}

\begin{document}

\title{
Fragmentation of massive dense cores down to $\lesssim1000$~AU:\\
Relation between fragmentation and density structure
}

\author{Aina Palau\altaffilmark{1},
Robert Estalella\altaffilmark{2},
Josep M. Girart\altaffilmark{1},
Asunci\'on Fuente\altaffilmark{3},
Francesco Fontani\altaffilmark{4},
Benoit Commer\c con\altaffilmark{5},
Gemma Busquet\altaffilmark{6},
Sylvain Bontemps\altaffilmark{7,8},
\'Alvaro S\'anchez-Monge\altaffilmark{4,9},
Luis A. Zapata\altaffilmark{10},
Qizhou Zhang\altaffilmark{11},
Patrick Hennebelle\altaffilmark{5},
James di Francesco\altaffilmark{12,13}
}
\altaffiltext{1}{Institut de Ci\`encies de l'Espai (CSIC-IEEC), Campus UAB -- Facultat de Ci\`encies, Torre C5 -- parell 2, 08193 Bellaterra, Catalunya, Spain}
\email{palau@ieec.uab.es}
\altaffiltext{2}{Departament d'Astronomia i Meteorologia (IEEC-UB), Institut de Ci\`encies del Cosmos, Universitat de Barcelona, Mart\'i i Franqu\`es, 1, 08028 Barcelona, Spain}
\altaffiltext{3}{Observatorio Astron\'omico Nacional, P.O. Box 112, 28803 Alcal\'a de Henares, Madrid, Spain}
\altaffiltext{4}{Osservatorio Astrofisico di Arcetri, INAF, Lago E. Fermi 5, 50125, Firenze, Italy}
\altaffiltext{5}{Laboratoire de Radioastronomie, UMR CNRS 8112, \'Ecole Normale Sup\'erieure et Observatoire de Paris, 24 rue Lhomond, 75231 Paris Cedex 05, France} 
\altaffiltext{6}{INAF-Istituto di Astrofisica e Planetologia Spaziali, Area di Recerca di Tor Vergata, Via Fosso Cavaliere 100, 00133, Roma, Italy}
\altaffiltext{7}{Universit\'e de Bordeaux, LAB, UMR 5804, 33270 Floirac, France}
\altaffiltext{8}{CNRS, LAB, UMR 5804, 33270 Floirac, France}
\altaffiltext{9}{I.\ Physikalisches Institut der Universit\"at zu K\"oln, Z\"ulpicher Stra{\ss}e 77, D-50937 K\"oln, Germany}
\altaffiltext{10}{Centro de Radioastronom\'ia y Astrof\'isica, Universidad Nacional Aut\'onoma de M\'exico, P.O. Box 3-72, 58090, Morelia, Michoac\'an, Mexico}
\altaffiltext{11}{Harvard-Smithsonian Center for Astrophysics, 60 Garden Street, Cambridge, MA 02138, USA}
\altaffiltext{12}{Department of Physics \& Astronomy, University of Victoria, PO Box 355, STN CSC, Victoria, BC, V8W 3P6, Canada}
\altaffiltext{13}{National Research Council Canada, 5071 West Saanich Road, Victoria, BC V9E 2E7, Canada}

\begin{abstract}
In order to shed light on the main physical processes controlling fragmentation of massive dense cores, we present a uniform study of the density structure of 19 massive dense cores, selected to be at similar evolutionary stages, for which their relative fragmentation level was assessed in a previous work.
We inferred the density structure of the 19 cores through a simultaneous fit of the radial intensity profiles at 450 and 850~\mum\ (or 1.2~mm in two cases) and the Spectral Energy Distribution, assuming spherical symmetry and that the density and temperature of the cores decrease with radius following power-laws. 
Even though the estimated fragmentation level is strictly speaking a lower limit, its relative value is significant and several trends could be explored with our data.
We find a weak (inverse) trend of fragmentation level and density power-law index, with steeper density profiles tending to show lower fragmentation, and vice versa. In addition, we find a trend of fragmentation increasing with density within a given radius, which arises from a combination of flat density profile and high central density and is consistent with Jeans fragmentation. We considered the effects of rotational-to-gravitational energy ratio, non-thermal velocity dispersion, and turbulence mode on the density structure of the cores, and found that compressive turbulence seems to yield higher central densities. Finally, a possible explanation for the origin of cores with concentrated density profiles, which are the cores showing no fragmentation, could be related with a strong magnetic field, consistent with the outcome of radiation magnetohydrodynamic simulations.
\end{abstract}

\keywords{
techniques: high angular resolution ---
stars: formation --- 
radio continuum: ISM ---
galaxies: star clusters
}

\section{Introduction \label{si}}

One unavoidable ingredient in the process of formation of intermediate/high-mass stars is that they are usually found associated with clusters of lower-mass stars (\eg\ Lada \& Lada 2003; Rivilla \et\ 2013).  However, the process yielding the formation of these clusters is not clear, as it remains ambiguous how a cloud core fragments to finally form a cluster. The observed star formation efficiency is by far much smaller than the predicted if gravity dominated the dynamics of the collapse (see \eg\ Kruijssen 2013 for a review), and the number of subcondensations typically embedded in massive dense cores is again smaller by one order of magnitude than the predicted from a pure Jeans fragmentation analysis (\eg\ Hennemann \et\ 2009; Zhang \et\ 2009, 2011; Bontemps \et\ 2010; Longmore \et\ 2011; Pillai \et\ 2011; Wang \et\ 2011; Lee \et\ 2013). Thus, a broad observational evidence indicates that additional physical ingredients competing with gravity must play an important role in the fragmentation process, such as angular momentum, stellar feedback, magnetic fields and turbulence. 

Numerical simulations are becoming crucial to establish the relative importance of each of these ingredients in the fragmentation process.
Simulations of fragmentation of rotating cores with different equations of state show that cores with some angular momentum easily fragment into multiple objects (\eg\ Boss \& Bodenheimer 1979; Bate \& Burkert 1997) and thus need of additional mechanisms to decrease the fragmentation down to the observed levels (see Goodwin et al. 2007 and references therein).
A crucial ingredient in this sense is the presence of a magnetic field. Studies on the effects of magnetic fields in the evolution of a collapsing and rotating core indicate that a strong magnetic field can suppress fragmentation as it constitutes an additional form of support and also via the so-called `magnetic braking' (\eg\ Hosking \& Whitworth 2004; Mellon \& Li 2008; Hennebelle \& Teyssier 2008), which is very effective in removing a significant part of the initial angular momentum through torsional Alfv\'en waves.
Similarly, radiation magnetohydrodynamic simulations of turbulent massive dense cores show that the combined effects of radiative feedback (\eg\ Krumholz \et\ 2007; Peters \et\ 2011) and magnetic field can sum to efficiently suppress fragmentation (\eg\ Hennebelle \et\ 2011, Commer\c con \et\ 2011; Myers \et\ 2013), yielding a better match with the observations (\eg\ Palau \et\ 2013). In addition, submillimeter observations show that the magnetic field is important in the dynamical evolution and fragmentation of massive dense cores (\eg\ Girart \et\ 2009, 2013; Hezareh \et\ 2013). 
Another ingredient which is probably playing a role in the fragmentation process is supersonic turbulence. Theoretical studies show that turbulence generates density structures due to supersonic shocks that compress and fragment the gas efficiently, especially for large Mach numbers (\eg\ Padoan \et\ 2001, Schmeja \& Klessen 2004; V\'azquez-Semadeni \et\ 2007). More recently it has been shown that the turbulence mode, solenoidal or compressive (\eg\ Federrath \et\ 2008), affects differently the cloud structure and fragmentation. 
Solenoidal turbulence corresponds to a null divergence of the driving force (divergence-free) and arises from galactic rotation and magneto-rotational instabilities, and seems to be at work in the Rosette and Polaris Flare clouds (\eg\ Hily-Blant \et\ 2008; Federrath \et\ 2010). 
Compressive turbulence corresponds to a null curl of the driving force (curl-free) and 
can be produced by expanding supernova shells or HII regions (\eg\ Federrath \et\ 2010), as seen in the Pipe Nebula and the Orion\,B cloud (Peretto \et\ 2012; Schneider \et\ 2013). 
Numerical simulations comparing both types of turbulence indicate that compressive turbulence promotes fragmentation and generates high density contrast structures, while solenoidal turbulence slows fragmentation down and keeps a smooth density structure in the cloud (\eg\ Federrath \et\ 2010; Girichidis \et\ 2011).
 
In addition to these physical agents opposing to gravity, the outcome of numerical simulations also depends on the geometry and physical properties of the collapsing core, in particular the density profile. Hydrodynamical simulations show that highly concentrated cores (\eg\ $n_\mathrm{H_2}\propto r^{-2}$) lead to the formation of one single massive star, while flat density profiles  ($n_\mathrm{H_2}\propto r^{-1}$ or flatter)  produce an important number of low-mass stars (\eg\ Myhill \& Kaula 1992; Burkert \et\ 1997; Girichidis \et\ 2011). 
This trend appears to be seen in high resolution observations of IRDC\,G28.34+0.06, where three massive dense cores with different density profiles show varying degrees of fragmentation (Zhang \et\ 2009; Wang \et\ 2011).
Finally, the substructure of molecular gas within a dense core is decreasing with time as the core turns gas into stars, as shown for example by Smith \et\ (2009).

From an observational point of view, most of the studies focusing on fragmentation of massive dense cores deal with single regions and/or do not reach high enough spatial scales to study individual fragments in the core ($\sim1000$--2000~AU; \eg\ Longmore \et\ 2011; Pillai \et\ 2011; Miettinen 2012; Beuther \et\ 2013; Kainulainen \et\ 2013; Takahashi \et\ 2013). 
Recently, Bontemps \et\ (2010) study a sample of six massive dense cores in Cygnus X and conducted a consistent analysis for all the regions, finding cores with a variety of fragmentation levels. A subsequent study by Palau \et\ (2013) focused on a list of 18 massive dense cores (including Bontemps \et\ 2010 sources), and observed with similar spatial resolution ($\sim1000$~AU), sensitivity (down to $\sim0.3$~\mo), and field of view ($\sim0.1$~pc of diameter), again reveals different fragmentation levels, with about $\sim30$\% of the regions showing no fragmentation at all.
Palau \et\ (2013) study possible correlations between the fragmentation level and properties of the cores such as mass, average density, evolutionary indicators, rotational-to-gravitational energy and turbulent velocity dispersion (although these last two properties should be studied in a larger sample), and find no clear trend of fragmentation level with any of these core properties. In that work, two main properties of the cores remained to be studied in relation to fragmentation, mainly the magnetic field and the density structure. While the magnetic field cannot currently be assessed in this sample in a uniform way, the density structure has been studied in roughly half of the sample (\eg\ Beuther \et\ 2002a; Mueller \et\ 2002; Crimier \et\ 2010; van der Tak \et\ 2013), but the methodology used is different for each work, casting some doubt on the validity of any comparison. 
Here we present a uniform study of the density structure of each core in the sample of Palau \et\ (2013), aiming at accurately assessing its impact on the fragmentation level.

In Section~2 we present the sample of cores and the data used to extract the submillimeter intensity profiles and the Spectral Energy Distributions (SEDs), and we also present the envelope model used to fit these data simultaneously for each core; in Section~3, we present the results of our fits; in Section~4, we discuss the possible trend between fragmentation level and density structure. Finally, in Section~5 we summarize our main conclusions.

\section{Data \label{sobs} and Methodology}

\subsection{Data}

We aim at studying the density structure of the 18 massive dense cores presented in Palau \et\ (2013) plus DR21-OH, a massive dense core for which fragmentation has been recently studied at similar mass sensitivities and spatial resolution by Girart \et\ (2013). Among the total number of 19 massive dense cores, 16 of them are identified as submillimeter sources in the catalog of Di Francesco \et\ (2008) and thus have James Clerk Maxwell Telescope (JCMT\footnote{The James Clerk Maxwell Telescope is operated by the Joint Astronomy Centre on behalf of the Science and Technology Facilities Council of the United Kingdom, the Netherlands Organisation for Scientific Research, and the National Research Council of Canada.}) 850 and 450~\mum\ data available (see Di Francesco \et\ 2008 for a detailed description on the observations). For the three cores not included in Di Francesco \et\ catalog, IRAS\,22198+6336, CygX-N12 and CygX-N63 (following labeling of Bontemps \et\ 2010), we used the submillimeter data of Jenness \et\ (1995), who observed also the 450 and 850~\mum\ emission with the JCMT, and the IRAM\,30m 1.2~mm data from Motte \et\ (2007). The beam (main + error) of the JCMT data was taken from Di Francesco \et\ (2008), and the beam (main + error) of the IRAM\,30m telescope at 1.2~mm was taken from Greve \et\ (1998). Three regions for which the 450~\mum\ data from Di Francesco \et\ (2008) were not usable were downloaded from the JCMT archive (IC1396N: m98bu24; AFGL5142: m96bc57; W3IRS5: m02ac32). The single-dish images at 450 and 850~\mum\ (1.2~mm) for the 19 cores, together with the interferometric images showing the fragmentation in each core, are shown in the Appendix (Figs. \ref{fobs1}--13).

\begin{figure}
\begin{center}
\begin{tabular}[b]{c}
    \epsfig{file=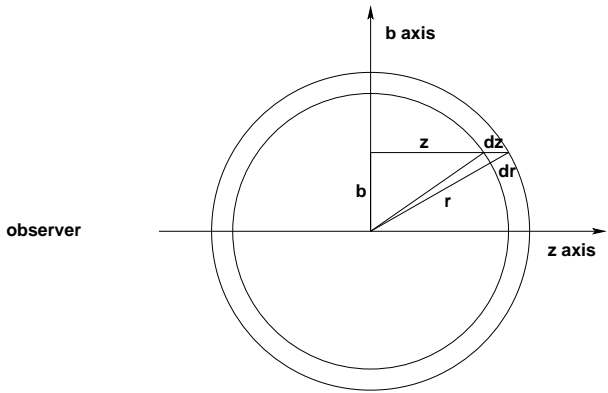, width=8cm, angle=0}\\
\end{tabular}
\caption{Sketch of the coordinate system used in the model described in Section~2.2.}
\label{fgeom}
\end{center}
\end{figure}

\begin{figure*}
\begin{center}
\begin{tabular}[b]{cc}
    \epsfig{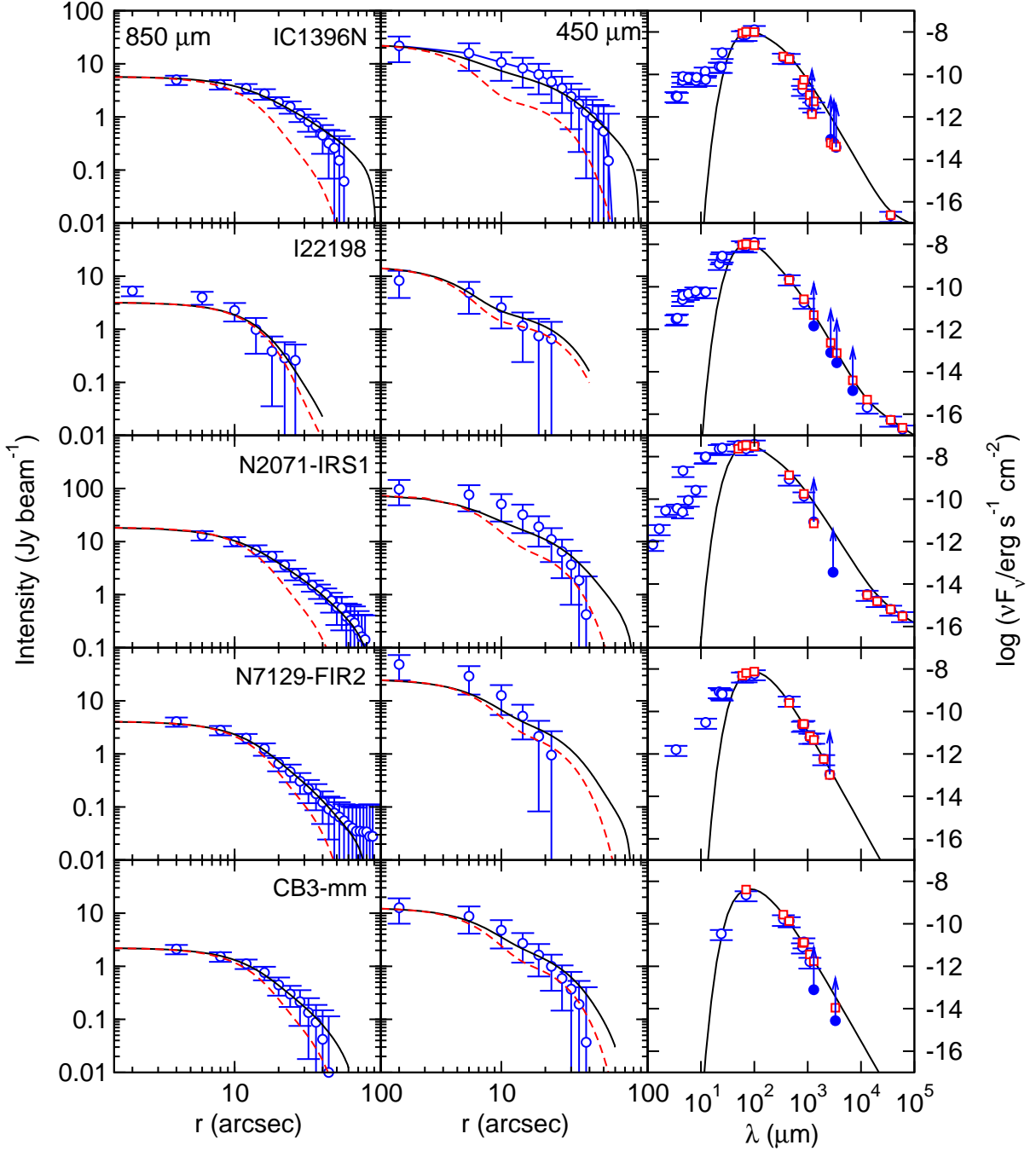}\\
\end{tabular}
\caption{Best fits for 5 regions of the sample (see Table~1 for the exact fitted parameters). Each row corresponds to one core, the left (middle) panel shows the radial intensity profile at 850 (450)~\mum, with the empty blue circles corresponding to the data, the black solid line corresponding to the model, and the dashed red line showing the beam profile; panels on the right show the SED, with blue empty circles indicating the observed fluxes, the blue full circles indicating observed lower limits, the black solid line showing the model for a fixed aperture, and the red squares corresponding to the model for the same aperture where each flux was measured.
}
\label{ffit1}
\end{center}
\end{figure*}

\begin{figure*}
\begin{center}
\begin{tabular}[b]{cc}
    \epsfig{file=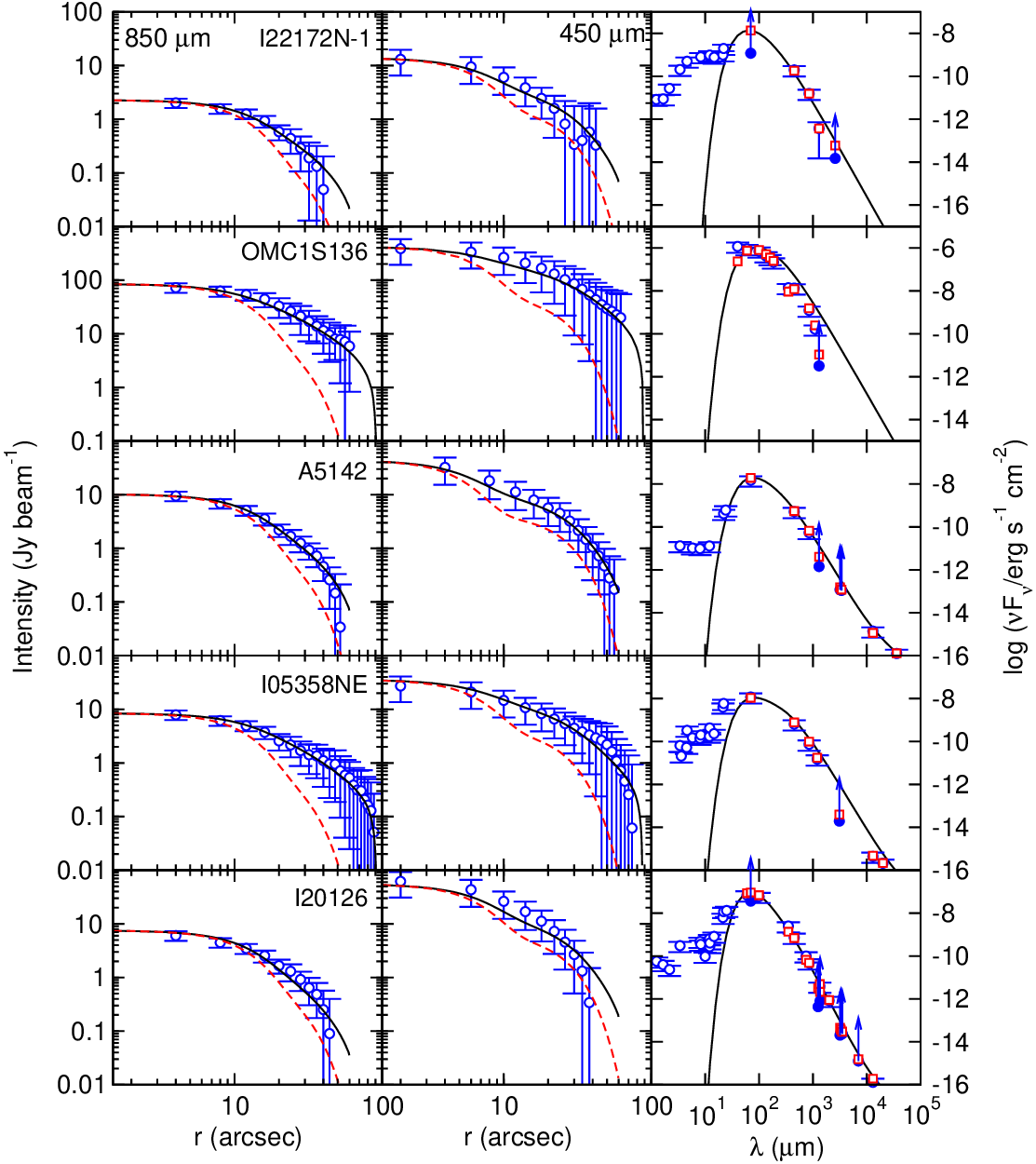, width=16cm, angle=0}\\
\end{tabular}
\caption{Same as Fig.~\ref{ffit1} for 5 different cores (Table~1).
}
\label{ffit2}
\end{center}
\end{figure*}

\begin{figure*}
\begin{center}
\begin{tabular}[b]{cc}
    \epsfig{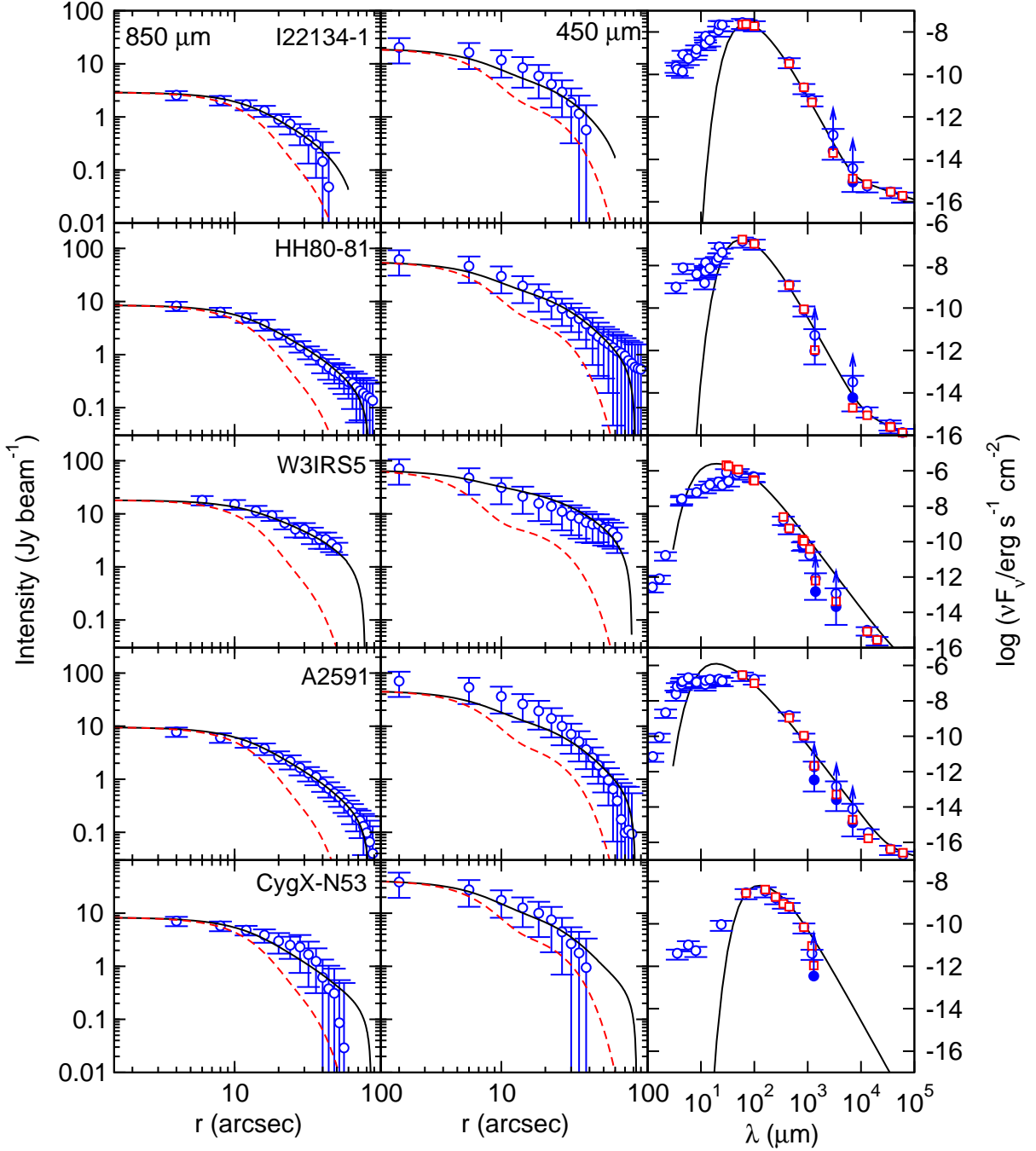}\\
\end{tabular}
\caption{Same as Fig.~\ref{ffit1} for 5 different cores  (Table~1).
}
\label{ffit3}
\end{center}
\end{figure*}

\begin{figure*}
\begin{center}
\begin{tabular}[b]{cc}
    \epsfig{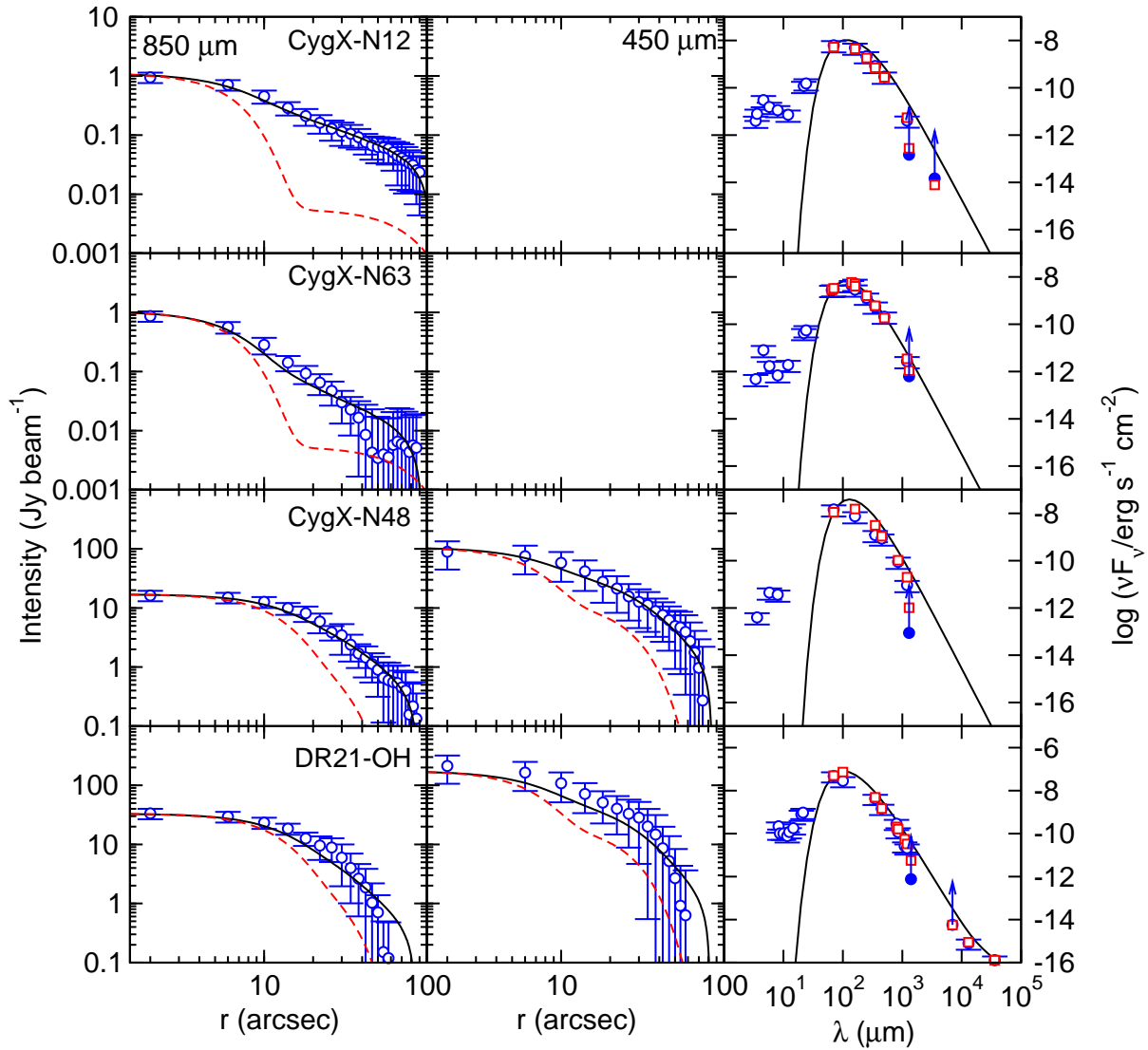}\\
\end{tabular}
\caption{Same as Fig.~\ref{ffit1} for the last 4 cores of Table~1. For CygX-N12 and CygX-N63 the profiles were extracted from the 1.2~mm images of Motte \et\ (2007).
}
\label{ffit4}
\end{center}
\end{figure*}

In order to characterize the spatial structure of the massive dense cores, we computed the circularly averaged radial intensity profile, in rings of $4''$
width, as a function of the projected distance from the core center at both 850 and 450~\mum\ when possible, and applied the model described in Fig.~\ref{fgeom} and Section~2.2. The intensity profiles at 850 and 450~\mum\ are shown in Figs.~\ref{ffit1}--\ref{ffit4}. As can be seen in the figures, the emission is partially resolved for all the cores, revealing the existence of an envelope of decreasing intensity with radius, even at small distances ($\la10''$) from the core center.
In some cases, the radial profile was calculated in particular angular `sectors' centered on the core peak, to avoid strong negative emission in the maps (NGC7129-FIRS2 at 450~\mum),
or contamination from a nearby strong core (W3IRS5) or from the DR21 ridge (CygX-N48, CygX-N53, DR21-OH). For these cases we did not use the values of the flux density at 850 and 450~\mum\ listed in Di Francesco's catalog, but remeasured the flux densities to assure that we were not including emission from the nearby cores/ridge. For the case of W3IRS5, the sector was taken in the north-south direction also to avoid possible contamination of free-free emission from strong centimeter sources\footnote{Although Richardson \et\ (1989a) estimated the contamination from free-free emission to be $<25$\% at 1.1mm, and only $<3$\% at 800~\mum, the centimeter sources A and B reported by Tieftrunk et al. (1997) are spatially overlapping the core emission and we prefer to avoid them.}. For the case of CygX-N48, we extracted the intensity profile avoiding the contribution from the north, where the strong DR21-OH core lies; for DR21-OH and CygX-N53 we extracted the profile in a sector with no restrictions in the east-west direction, but allowing it to extend $\sim \pm25''$ in the north-south direction, 
which corresponds to the approximate width of the DR21 filament 
in the east-west direction.

The SED of each massive dense core was built from centimeter wavelengths up to 50--60~\mum, as this is approximately the wavelength where envelope and disk contributions start to be comparable 
(\eg\ Crimier \et\ 2010; Commer\c con \et\ 2012), 
and here we only model the envelope contribution. To build the SEDs we used the data compiled in Palau \et\ (2013), which included flux densities at different frequencies from IRAS, JCMT/SCUBA, AKARI and IRAM\,30m data when available. In addition, we measured the flux densities at 70~\mum\ of cores with available Spitzer/MIPS2 data. Flux densities obtained in heavily saturated Spitzer/MIPS2 data were considered as lower limits. We also considered as lower limits the millimeter/submillimeter flux densities measured with an interferometer, as part of the core emission is filtered out and we did not take this into account in our model. For OMC-1S we recompiled the SED because the SED in Palau \et\ (2013) was built for the source 136$-355$ (about $\sim12''$ north of the submillimeter peak), while now we have considered the fluxes associated with the submillimeter peak, as we are modeling the submillimeter profiles. In addition, we used the FIR/submillimeter/centimeter fluxes from the literature\footnote{ 
IC1396N: Saraceno \et\ (1996), Beltr\'an \et\ (2002b), Codella \et\ (2001);
I22198: S\'anchez-Monge (2011);
NGC\,2071: Butner et al. (1990); 
NGC7129: Eiroa et al. (1998), Fuente et al. (2001), Crimier \et\ (2010);
CB3: Launhardt, Ward-Thompson \& Henning (1997); 
OMC1-S: Jaffe \et\ (1984), Lis \et\ (1998), Goldsmith, Bergin \& Lis (1997), Zapata \et\ (2004a, b);
AFGL5142: Zhang \et\ (2007), Hunter \et\ (1999), S\'anchez-Monge (2011);
I05358NE, I22134: Beuther \et\ (2002, 2007), S\'anchez-Monge (2011);
I20126: Cesaroni \et\ (1997, 1999, 2005), Hofner \et\ (1999), S\'anchez-Monge (2011);
I22134: S\'anchez-Monge (2011);
HH80: Fern\'andez-L\'opez \et\ (2011);
W3IRS5: Campbell et al. (1995), Richardson \et\ (1989a); 
AFGL\,2591: van der Tak \et\ (1999), van der Tak \& Menten (2005);
CygX-N53, N12, N63 and N48: Duarte-Cabral \et\ (2013);
DR21-OH: Harvey \et\ (1986), Gear \et\ (1988), Richardson \et\ (1989b), Motte \et\ (2007), Araya \et\ (2009), Zapata \et\ (2012).
}. 
The apertures where fluxes were assessed were figured out from the literature or adopted as reported in the used catalogs and references (see above), and for peak intensities we took the beam as aperture radius.

\subsection{Model}

We modeled the thermal emission of the cores at millimeter and submillimeter wavelengths with a spherically symmetric envelope of gas and dust, with density and temperature decreasing with radius.
The density and temperature were assumed to be power-laws of radius, with indices
$p$ and $q$, $\rho=\rho_0(r/r_0)^{-p}$ and $T=T_0(r/r_0)^{-q}$.
For the dust opacity law we assumed that the dust opacity is a power-law of
frequency with index $\beta$, $\kappa_\nu=\kappa_0(\nu/\nu_0)^\beta$, where
$\nu_0$ is an arbitrary reference frequency. The values used for the dust
opacity were those of Ossenkopf \& Henning (1994), for thin ice mantles at a gas density of $10^6$
cm$^{-3}$,  i.e.\ $\kappa_0=0.008991$ cm$^2$ g$^{-1}$ 
for $\nu_0=230$~GHz, being $\beta$ a free parameter of the model.

In the case of optically-thin  dust emission, in the Rayleigh-Jeans
approximation, and for an infinite radius envelope, the intensity as a function
of the projected distance to the source center, $I_\nu(b)$, can be derived
analytically (see, for instance, Beltr\'an \et\ 2002a), and is a power-law of the
projected distance to the envelope center $b$, $I_\nu(b)\propto b^{1-(p+q)}$,
with index  $1-(p+q)$.

In our model, however, none of the above approximations (optically thin, Rayleigh-Jeans) were made, and there is
no analytical expression for the intensity profile. In general, the contribution
to the radiated intensity from an element of the envelope of geometrical depth
$dz$, at a radius $r$ from the center (see Fig.\ \ref{fgeom}) is
given by
\begin{equation}\label{ethick}
dI_\nu = \frac{dI_\nu(r)}{dz}\,dz = 
\frac{2k\nu^2}{c^2}\,J_\nu[T(r)]\,e^{-\tau_\nu(b,z)} \rho(r)\,\kappa_\nu\,dz,
\end{equation}
where
\begin{equation}\label{eqtau}
\tau_\nu(b,z) = \int_{-\infty}^z \rho(r)\,\kappa_\nu\,dz',
\end{equation}
with $r=\sqrt{b^2+z^2}$, is the optical depth from the observer to the element 
of the envelope at a projected distance $b$ and position $z$ along the line of 
sight (see Fig.\ \ref{fgeom}), and
$J_\nu(T)$ is the Planck function in units of temperature,
\begin{equation}
J_\nu(T) = \frac{h\nu/k}{\exp(h\nu/kT)-1}.
\end{equation}
Then, the resulting intensity as a function of 
the projected distance to the source, $I_\nu(b)$, has to be evaluated as
\begin{equation}\label{eqib}
I_\nu(b) = \int_{-z_\mathrm{max}}^{+z_\mathrm{max}} \frac{dI_\nu(r)}{dz}\, dz,
\end{equation}
where $z_\mathrm{max}=\sqrt{b^2+{r_\mathrm{max}}^2}$ and $r_\mathrm{max}$ is the
maximum radius where the model is calculated.
The computational burden is low, since  both integrals, those of Eqs.\
\ref{eqtau} and \ref{eqib}, can be evaluated simultaneously in a single run.

Regarding the temperature distribution of the envelope, Chandler \et\ (1998) derive from the expressions of Scoville \et\ (1976) that, for the radiative heating of a dust cloud by a central luminous source, the temperature power-law index $q$ is a function of the dust emissivity index $\beta$, $q=2/(4+\beta)$. 
However, for the outer part of the envelope the external heating can be important, and a constant temperature of 10~K was adopted for radii larger than the radius where the temperature distribution of index $-q$ drops to 10~K. 
Regarding the density distribution, we adopted, as maximum radius of the envelope $r_\mathrm{max}$, the radius for which the envelope density dropped to a value comparable to that of the ambient gas of the intercore medium, taken to be 5000~cm$^{-3}$.

In summary, the four free parameters of the model are 
the dust emissivity index, $\beta$; 
the envelope temperature at the reference radius $r_0$, $T_0$;
the envelope density  at the reference radius $r_0$, $\rho_0$;
and the density power-law index, $p$.
The reference radius $r_0$ is arbitrary, and has been taken as $r_0=1000$ AU.

With this model we fitted the observed radial intensity profiles at 450 and 850
\mum\ or 1.2 mm, and the observed SED from
centimeter wavelengths up to 60~\mum\ simultaneously. The uncertainty adopted
for the intensity profiles was $20$\% at 850~\mum\ and 1.2~mm, and $50$\% at
450~\mum\ (Di Francesco \et\ 2008). The uncertainty of the flux
densities used to build the SED was adopted to be $50$\%.

In order to compare the model with the observed intensity profile at each
wavelength, we computed the intensity map from the model, and we
convolved it with the beam, given as the sum of two circular Gaussians (the main
and error beams). 
Finally, the circularly averaged profile was computed from
the convolved map.
We also simulated the effect of the finite chop throw used in the observations
to cancel the sky emission. Thus, for each map position we estimated the
off-source intensity as the average intensity at a chop-throw distance of
$120''$ from the map (on-source) position, and we computed the difference
between the on- and off-source intensities. From this, we computed the radial
profile corrected for chopping.

In order to compare the model with the observed SED, for each data point  with
wavelength higher than 60 $\mu$m we computed the model flux density by
integration of the model intensity profile inside the reported aperture used for
each measurement.  In addition, for points with frequency below 1000 GHz, we
added the free-free emission of an unresolved source with a flux density
extrapolated from the data points at centimeter wavelengths.

The fitting procedure was the sampling of the four-dimensional parameter space, using the same procedure as that described in Estalella \et\ (2012) and S\'anchez-Monge \et\ (2013). For every set of parameters, we computed the residual
$\chi^2$,
\begin{equation}\label{eqchi2}
\chi^2\equiv\sum_{i=1}^n
\left[
\frac{y_i^\mathrm{obs}-y_i^\mathrm{mod}(\beta, T_0, \rho_0, p)}{\sigma_i}
\right]^2,
\end{equation}
where the sum encompasses the points of the intensity profiles at 850 and 450
$\mu$m, and the points of the SED fitted by the model.
The parameter space was searched for the minimum value of $\chi^2$. 
For each source, the initial search ranges for the four parameters was
$\beta=1.5\pm1.5$,
$T_0=300\pm300$~K, 
$\rho_0=(1.0\pm1.0)\times10^{-16}$~g cm$^{-3}$, and
$p=1.5\pm1.0$.
The search range was reduced a factor of 0.8 around the best fit value of the
parameters found for each loop, consisting in 2000 samples of the parameter
space. The final best fit values were taken after 10 loops.
Once the best fit parameters were found, their uncertainties were estimated
through the increase in  $\chi^2$ (see Estalella \et\ 2012, S\'anchez-Monge \et\ 2013).
We assessed the quality of the fits through the value of the reduced chi-square,
\begin{equation}
\chi_r\equiv\sqrt{\frac{\chi^2_\mathrm{min}}{n-4}}.
\end{equation}
In general for all the sources we found fits with $\chi_r\simeq1$, or even less,
indicating that the errors of the data were slightly overestimated.
In Fig.~14 we show as example, the parameter space around the best
fit solution for the source IC\,1396N.

\begin{table*}
\caption{Best-fit parameters to the radial profiles and SED of the massive dense cores}
\begin{center}
{\small
\begin{tabular}{lcr cccc cccc}
\noalign{\smallskip}
\hline\noalign{\smallskip}
&D
&$L_\mathrm{bol}$\supa
&&
&$T_0$\supb
&$\rho_0$\supb
&
\\
ID-Source
&(kpc)
&(\lo)
&$N_\mathrm{mm}$\supa
&$\beta$\supb
&(K)
&(g cm$^{-3}$)
&$p$\supb
&$\chi_r$\supb
&$p_\mathrm{lit}$\supc
&Refs.\supc
\\
\noalign{\smallskip}
\hline\noalign{\smallskip}
1-IC1396N	      	&0.75	&290	&4	&$1.41\pm0.19$	&$43\pm4$	&$(2.8\pm0.4)\times10^{-17}$	&$1.62\pm0.08$	&0.580	&1.2	&1\\
2-I22198\supd        	&0.76	&340	&1.5	&$1.46\pm0.31$	&$43\pm5$	&$(2.8\pm1.0)\times10^{-17}$	&$2.45\pm0.16$	&0.885	&$-$	&$-$\\
3-NGC2071-IRS1 	&0.42	&440	&4	&$1.09\pm0.21$	&$40\pm3$	&$(6.1\pm1.0)\times10^{-17}$	&$1.83\pm0.09$	&0.534	&$-$	&$-$\\
4-NGC7129-FIRS2	&1.25	&460       	&1	&$1.55\pm0.28$	&$47\pm4$	&$(5.9\pm1.1)\times10^{-17}$	&$2.14\pm0.11$	&0.454   	&1.4	&1\\
5-CB3-mm		&2.50	&700	&2	&$1.42\pm0.24$	&$58\pm7$	&$(7.4\pm1.5)\times10^{-17}$	&$2.04\pm0.10$	&0.552   	&2.2	&1\\
6-I22172N-IRS1      	&2.40	&830     	&3	&$1.49\pm0.23$	&$75\pm10$	&$(3.7\pm0.7)\times10^{-17}$	&$1.89\pm0.08$	&0.283   	&$-$	&$-$\\
7-OMC-1S        	&0.45	&2000  	&9	&$1.42\pm0.20$	&$86\pm9$	&$(8.8\pm1.8)\times10^{-17}$	&$1.56\pm0.10$	&0.319	&??	&2\\
8-AFGL\,5142		&1.80	&2200     	&7	&$1.25\pm0.20$	&$70\pm7$	&$(1.8\pm0.2)\times10^{-16}$	&$2.00\pm0.05$	&0.361 	&$-$	&$-$\\
9-I05358+3543NE	&1.80	&3100   	&4	&$1.28\pm0.18$	&$62\pm6$	&$(6.4\pm1.0)\times10^{-17}$	&$1.55\pm0.05$	&0.229   	&$>0.8$,1.5	&3,7\\
10-I20126+4104	&1.64	&8900	&1	&$1.82\pm0.24$	&$86\pm9$	&$(8.4\pm1.6)\times10^{-17}$	&$2.21\pm0.11$	&0.607   	&1.6,1.8,2.2&3,4,5\\
11-I22134-IRS1      	&2.60	&11800  	&3.5	&$1.70\pm0.19$	&$82\pm8$	&$(3.1\pm0.5)\times10^{-17}$	&$1.76\pm0.06$	&0.477     &1.3	&3\\ 
12-HH80-81         	&1.70	&21900	&3	&$1.56\pm0.14$	&$108\pm10$	&$(4.2\pm0.6)\times10^{-17}$	&$1.78\pm0.04$	&0.473 	&$-$	&$-$\\
13-W3IRS5	    	&1.95	&140000	&3.5	&$1.04\pm0.12$	&$260\pm30$	&$(2.4\pm0.3)\times10^{-17}$	&$1.46\pm0.04$	&0.602	&1.5,1.4	&4,7\\
14-AFGL\,2591		&3.00	&190000	&1.5	&$0.96\pm0.12$	&$250\pm20$	&$(5.4\pm0.7)\times10^{-17}$	&$1.80\pm0.03$	&0.549	&1.0,2.0,1.0&4,6,7\\
\hline
15-CygX-N53		&1.40	&300	&6	&$1.55\pm0.22$	&$45\pm4$	&$(9.7\pm1.8)\times10^{-17}$	&$1.76\pm0.07$	&0.487	&$-$	&$-$\\
16-CygX-N12\supe	&1.40	&320	&2.5	&$1.75\pm0.26$	&$50\pm4$	&$(3.3\pm0.3)\times10^{-17}$	&$1.52\pm0.03$	&0.376	&$-$	&$-$\\
17-CygX-N63\supe	&1.40	&470	&1	&$1.80\pm0.33$	&$45\pm3$	&$(6.5\pm1.1)\times10^{-17}$	&$2.03\pm0.07$	&0.570	&$-$	&$-$\\
18-CygX-N48		&1.40	&4400	&5	&$1.88\pm0.18$	&$58\pm5$	&$(1.0\pm0.2)\times10^{-16}$	&$1.71\pm0.05$	&0.459	&$-$	&$-$\\
19-DR21-OH		&1.40	&10000	&11	&$1.60\pm0.26$	&$73\pm7$	&$(3.1\pm0.6)\times10^{-16}$	&$1.98\pm0.08$	&0.808	&1.8,1.4	&4,7\\
\hline
\end{tabular}
\begin{list}{}{}
\item[$^\mathrm{a}$] \Lbol\ taken from Palau \et\ (2013). \Nmm\ is the fragmentation level, taken from Palau \et\ (2013) and assessed by estimating the number of millimeter sources within a field of view of 0.1~pc of diameter as detected with a millimeter interferometer with a mass sensitivity of $\sim0.3$~\mo\ and a spatial resolution of $\sim1000$~AU. For the four Cygnus\,X cores, \Nmm\ has been revised with respect to the values given in Palau \et\ (2013).
\item[$^\mathrm{b}$] Free parameter fitted by the model: $\beta$ is the dust emissivity index; $T_0$ and $\rho_0$ are the temperature and density at the reference radius, 1000 AU; $p$ is  the density power law index; $\chi_\mathrm{r}$ is the reduced $\chi$ as defined in equation (6).
\item[$^\mathrm{c}$] Density power-law index from different works in the literature. Refs: 1: Crimier \et\ (2010);  2: Mezger \et\ (1990); 3: Beuther \et\ (2002); 4: van der Tak \et\ (2000); 5: Shinnaga \et\ (2008); 6: Mueller \et\ (2002); 7: van der Tak \et\ (2013). For I22134, the reported value is taken from Beuther \et\ (2002) who fitted two power-law indices with a break at $32''$, and we give here the outer power-law index. To convert from the reported intensity power-law index to the density power law index, Beuther \et\ (2002) adopt a temperature power-law index of 0.4. Similar for Shinnaga \et\ (2008), for which the power-law break is taken at $12''$.
\item[$^\mathrm{d}$] Digitized data from Jenness \et\ (1995). 
\item[$^\mathrm{e}$] Sources for which only the radial intensity profile at 1.2~mm was fitted. For CygX-N63, \Nmm\ was inferred from new Plateau de Bure observations in AB configuration (S. Bontemps, priv. communication).
\end{list}
}
\end{center}
\label{tfit1}
\end{table*}

\section{Results \label{sres}}

Figures~\ref{ffit1}-\ref{ffit3} present the best simultaneous fit to the 850 and 450~\mum\ (or 1.2~mm) radial intensity profiles and the SED for the 19 massive dense cores. A summary of the fitted parameters is presented in Table~\ref{tfit1}. The four fitted parameters span a wide range of values. The dust emissivity index ranges from 1.0 to 1.9, with a mean value of 1.5. This is comparable to dust emissivity indices measured in high-mass envelopes (\eg\ Mueller \et\ 2002; Hill \et\ 2006 for temperatures between 30 and 50~K), 
and corresponds to temperature power-law indices ranging from 0.34 to 0.40 (mean value: 0.37, consistent with Emerson 1988). The fitted values for the temperature at 1000~AU range from 40 to 260~K, with a mean value of 83~K, and correlate with the bolometric luminosity
\footnote{The temperature at 1000~AU and the bolometric luminosity correlate with a slope of +0.24, following approximately the Stephan-Boltzman law (of slope +0.25, Table~1).}. 
The range of densities at 1000~AU is (3--30)$\times10^{-17}$~g\,cm$^{-3}$, with a mean value of $\sim8\times10^{-17}$~g\,cm$^{-3}$. Finally, the index of the density power law ranges from 1.5 to 2.4, with a mean value of 1.85, similar to the results found in previous works focused on massive dense cores (\eg\ Pirogov \et\ 2009). The cores with steeper density profiles are I22198, NGC\,7129, and I20126, while the cores presenting the flattest profiles are OMC-1S, I05358NE, W3IRS5 and CygX-N12.

Once the model was fitted to the data, we inferred different quantities, which are given in Table~2. The radius where the density reaches $\sim5000$~\cmt, considered as the ambient density, ranges from 0.1 to 1.0~pc. The regions with smallest radii were I22198, NGC7129-1 and CygX-N63, in part due to the concentrated density profile inferred for these regions, of $p\gtrsim2$. On the other hand, the cores with largest radii are I05358NE and CygX-N48. We estimated the `observed mass' of the core
by integrating the density profile up to the observed radius of the core, which ranged from 5 to 1000~\mo. 
In addition, we estimated the surface density (\Surfd) and the density (\nradius) of each core within a radius of 0.05~pc, as this is the radius where we assessed the fragmentation level. We found that 
\Surfd\ ranged from $\sim0.1$ up to 1.8~g\,cm$^{-2}$, with a mean value of 0.6~g\,cm$^{-2}$. 
Finally, the radii where the temperature falls down to 10~K ranged from 0.2 to 18~pc, being the cores with higher temperature (at reference radius) those where this radius is larger.

\begin{table*}
\caption{Inferred and compiled properties of the dense cores}
\begin{center}
{\small
\begin{tabular}{lccc cccc cccc cc}
\noalign{\smallskip}
\hline\noalign{\smallskip}
&
&$r_\mathrm{10\,K}$\supa
&$r_\mathrm{max}$\supa
&$M_\mathrm{obs}$\supa
&$\Sigma_\mathrm{0.05\,pc}$\supa
&$n_\mathrm{0.05\,pc}$\supa
&
&$\sigma_\mathrm{noth-init}$\supb
&$\sigma_\mathrm{noth-feedb}$\supb
&$M_\mathrm{vir}$\supb
&
&turb.
\\
ID-Source
&$q$\supa
&(pc)
&(pc)
&(\mo)
&(g\,cm$^{-2}$)
&($\times 10^5$\,cm$^{-3}$)
&$\beta_\mathrm{rot}$\supb
&(\kms)
&(\kms)
&(\mo)
&$\alpha_\mathrm{vir}$\supb
&mod.\supc
&R.\supc
\\
\noalign{\smallskip}
\hline\noalign{\smallskip}
1-IC1396N       	  	&0.37	&0.26	&0.38	&78		&0.28	&3.6		&0.016	&$-$		&$-$		&$-$		&$-$		&c?	&1\\
2-I22198        		&0.37	&0.27	&0.09	&5		&0.10	&1.3		&0.003	&0.51	&0.46	&10		&2.4		&?	&$-$\\
3-N2071-1		&0.39	&0.17	&0.36	&80		&0.45	&5.7	   	&0.066	&0.42	&0.72	&20		&1.2		&c	&2\\
4-N7129-2		&0.36	&0.36 	&0.19	&81		&0.29	&3.6	    	&$-$		&$-$		&$-$		&$-$		&$-$		&?	&$-$\\
5-CB3-mm		&0.37	&0.56	&0.25	&169	&0.41	&5.2		&$-$		&$-$		&$-$		&$-$		&$-$		&c?	&3\\
6-I22172N      		&0.36	&1.23	&0.24	&119	&0.25	&3.2      	&0.062	&0.59	&0.58	&16		&1.8		&c?	&4\\
7-OMC-1S      		&0.37	&1.64	&0.96	&158	&1.00	&13 		&0.020	&$-$		&1.15	&51		&1.3		&c	&5\\
8-A5142         		&0.38	&0.81	&0.42	&356	&1.04	&13	  	&0.011	&0.63	&1.61	&94		&2.4		&c?	&6\\
9-I05358NE		&0.38	&0.60	&0.79	&1480	&0.72	&9.1	     	&0.005	&0.51	&0.72	&21		&0.8		&c?	&7\\
10-I20126			&0.34	&2.51	&0.20	&68		&0.38	&4.8	    	&0.004	&0.38	&2.00	&144	&10		&?	&$-$\\
11-I22134      		&0.35	&1.98	&0.29	&222	&0.26	&3.2	       	&0.072	&0.33	&0.40	&10		&1.0		&?	&$-$\\ 
12-HH80-81         	&0.36 	&3.64	&0.32	&333	&0.33	&4.2	      	&0.041	&0.68	&0.97	&38		&3.2		&?	&$-$\\ 	
13-W3IRS5          	&0.40	&18.5	&0.56 	&971 	&0.32	&4.0		&0.021	&$-$		&2.16	&175	&15		&?	&$-$\\
14-A2591			&0.40	&13.9	&0.36	&784	&0.41	&5.2		&0.013	&$-$		&1.56	&97		&6.1		&?	&$-$\\
\hline	
15-Cyg-N53		&0.36	&0.32	&0.56	&675	&0.81	&10		&$-$		&$-$		&0.41	&8		&0.3		&c	&8\\
16-Cyg-N12		&0.35	&0.51	&0.57	&622	&0.39	&5.0		&$-$		&$-$		&$-$		&$-$		&$-$		&c	&9\\
17-Cyg-N63		&0.34	&0.39	&0.24	&160	&0.37	&4.6		&$-$		&$-$		&$-$		&$-$		&$-$		&?	&$-$\\
18-Cyg-N48		&0.34	&0.86	&0.67	&865	&0.94	&12		&$-$		&0.61	&1.06	&42		&1.2		&c	&8\\
19-DR21-OH		&0.36	&1.26	&0.58	&526	&1.83	&23		&0.240	&0.61	&1.48	&81		&1.2		&c	&8\\
\hline
\end{tabular}
\begin{list}{}{}
\item[$^\mathrm{a}$] Parameters inferred (not fitted) from the modeling. $q$ is the temperature power-law index, and $r_\mathrm{10\,K}$ is the radius of the core where the temperature has dropped down to $\sim10$~K; $r_\mathrm{max}$ is the radius at the assumed `ambient' density of 5000~\cmt; $M_\mathrm{obs}$ is the mass computed analytically from the model integrating until the radius where the density profile could be measured for each source. $\Sigma_\mathrm{0.05\,pc}$ and \nradius\ are the surface density and density inside a region of 0.05~pc of radius.
\item[$^\mathrm{b}$] \betar\ is the rotational-to-gravitational energy ratio as in Palau \et\ (2013) and adding the new measured values from \nh(1,1) for OMC-1S (Wiseman \& Ho 1998) and DR21-OH (Mangum \et\ 1992). 
$\sigma_\mathrm{noth-init}$ is the non-thermal velocity dispersion for a quiescent nearby core (thus a first approach to the initial velocity dispersion), as in Palau \et\ (2013) and adding a new value for CygX-N48 and DR21-OH (Mangum \et\ 1992). $\sigma_\mathrm{noth-init}$ was estimated assuming a temperature of 20~K.
$\sigma_\mathrm{noth-feedb}$ is the non-thermal velocity dispersion for the studied core (thus including outflow feedback from the nascent protostars) from data in the literature (see Section~\ref{sdaffectingdens} for references). $\sigma_\mathrm{noth-feedb}$ was estimated using the temperature at radius 0.05~pc as inferred from the modeled temperature profile (Tables~1 and 2).
$M_\mathrm{vir}$ and $\alpha_\mathrm{vir}=M_\mathrm{vir}/M_\mathrm{0.05pc}$ (with $M_\mathrm{0.05pc}$ being the mass within a radius of 0.05~pc) are assessed using the non-thermal velocity dispersion including feedback from outflows. 
\item[$^\mathrm{c}$] `Turb. mode' refers to tentative turbulence mode according to works in the literature: `c' indicates that a detailed study has been done comparing the properties of the region with simulations or studying the Probability Density Function; `c?' indicates that the only hints of possible compressive turbulence are that the region lies at the border of an expanding bubble/HII region. References about turbulence mode: 1: Beltr\'an \et\ (2002b); 2; Schneider \et\ (2013); 3: Sakai \et\ (2013);
4: Liu \et\ (2012); 5: L\'opez-Sepulcre \et\ (2013); 6: Hunter \et\ (1995); 7: Leurini \et\ (2007); 8: Schneider \et\ (2011); 9: Schneider \et\ (2006).
\end{list}
}
\end{center}
\label{tfit2}
\end{table*}

\section{Discussion}

\subsection{Comparison of density structure with previous works}

We compared our inferred density power-law indices to the results of previous works where a similar methodology was applied for some of the cores of our sample. In the two last columns of Table~\ref{tfit1} we list $p_\mathrm{lit}$, the density power-law index from the literature, together with the corresponding references.
Crimier \et\ (2010), van der Tak \et\  (2000, 2013) and Mueller \et\ (2002) fit one single power-law to the radial intensity profiles, being comparable to the work presented here, while Beuther \et\ (2002) used two power-laws with a break at $32''$. 
Within the regions fitted by the same authors the tendency of flat and concentrated profiles is consistent with the results found in the present work (for example, IC1396N is flatter than NGC7129-FIRS2 and CB3-mm as in Crimier \et\ 2010; or I20126 is more concentrated than I22134 and I05358NE as in Beuther \et\ 2002). 
However, the absolute value of the density power-law indices cannot be directly compared between different works because of the different methods applied, which yield differences of 0.2--0.4 in the density power-law index.
The most important difference between previous works and this work is the way of determining the temperature profile. While Mueller \et\ (2002), Crimier \et\ (2010), and van der Tak \et\ (2013) calculate the temperature profile self-consistently through radiative transfer codes, we assume that the temperature profile is related to the dust emissivity index (Section 2.2) and leave the temperature at a given radius as a free parameter. This is also different to Beuther \et\ (2002) work, who assumed a fixed temperature power-law index of 0.4. Our method finds in general flatter temperature profiles compared to Crimier \et\ (2010) and van der Tak \et\ (2013), yielding higher density power-law indices. On the other hand, our method is perfectly consistent with the results of Mueller \et\ (2002) for AFGL\,2591. 
Therefore, once the different methodologies are taken into account, our methodology yields consistent results with previous works, and allows to fairly compare the fitted parameters from one region to the other.

\begin{figure}
\begin{center}
\begin{tabular}[b]{c}
    \epsfig{file=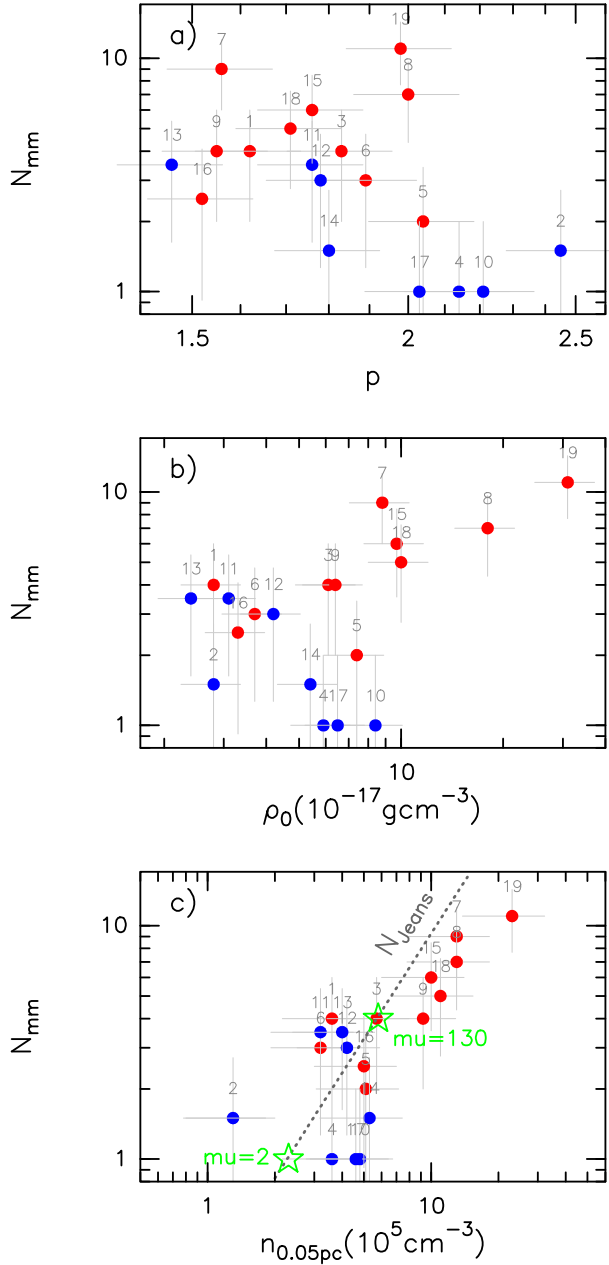, width=8.0cm, angle=0}\\
\end{tabular}
\caption{Relation between the fitted parameter $p$ (density power-law index, panel "a"), $\rho_0$ (density at a radius of 1000~AU, panel "b"), and \nradius\ (density within 0.05~pc of radius, panel "c") with \Nmm\ (fragmentation level, see Table~1). In the last panel, the Jeans number is calculated as $M_\mathrm{0.05pc}\times CFE/M_\mathrm{Jeans}$, assuming a temperature of 20~K and a Core Formation Efficiency (CFE) of 17\% (see main text), and the green stars indicate the measurements in the simulations of Commer\c con \et\ (2011) for the $\mu=2$ (strongly magnetized) and $\mu=130$ (weakly magnetized) cases.
In all panels, red dots correspond to cores with hints of compressive turbulence, and blue dots correspond to cores with no assessment of the turbulence mode in the literature (see Table~\ref{tfit2} and Section~4.4).
}
\label{fcorr}
\end{center}
\end{figure}

\subsection{Uncertainty in the fragmentation level and possible biases in the sample}

Before comparing the inferred density structure for each massive dense core (estimated through the density power-law index $p$), to the fragmentation level (estimated through the \Nmm\ parameter, see Table~\ref{tfit1} and Palau \et\ 2013), we discuss here the main uncertainties in the assessment of these properties and possible biases. First of all we stress that the determination of both $p$ and \Nmm\ has been carried out applying the same methodology and criteria to all the cores and thus, although the absolute values might be subject to discussion, the relative value between the regions is meaningful and valid for a comparison.
In particular, \Nmm\ is a lower limit to the total number of fragments because of flux missed by the interferometers, and because we are missing objects emitting at wavelengths different to millimeter. However, the sample presented in Palau \et\ (2013) was selected applying strict criteria for the filtered emission and $uv$-coverage
(see column 8 of Table~3 of Palau \et\ 2013), constraining this study to the most compact fragments ($\la5000$~AU). 
Since larger non-centrally peaked structures might be transient, this is not likely affecting our estimate of the fragmentation level. 
In addition, the number of infrared sources were compared to \Nmm\ and were found to be correlated, indicating that \Nmm\ is systematically underestimating the `real' fragmentation level of the core by only a factor of $\sim2$. 
Finally, Palau \et\ (2013) also used radiation MHD simulations to show that \Nmm\ is sensitive to the original fragmentation of the core.
Altogether suggests that \Nmm\ is already a good indicator of the fragmentation level if used to compare the different regions and not as an absolute value.

A possible bias could come from the different distances in the sample. However, for 16 out of 19 cores distances range only from 0.8 to 2.5 kpc, and no trend of \Nmm, $p$, or any of the fitted parameters vs distance is found in our data (Table~1), indicating that this is not affecting our conclusions.
Another bias could arise from different evolutionary stages of the cores of our sample, as simulations show that the degree of substructure in a cloud decreases with age (\eg\ Smith \et\ 2009). 
Our sample  is not likely dramatically affected by this because the targets were selected to be in similar evolutionary stages: all targets harbor an intermediate/high-mass protostar detected in the far-infrared, are associated with a strong and compact millimeter core, are associated with molecular outflows, and have not developed an (ultra-)compact HII region yet. The small differences in evolutionary stage of the cores of this sample were also studied in Palau \et\ (2013) by estimating evolutionary indicators, and no relation between \Nmm\ and these indicators was found (see e.g., Fig. 8B of Palau \et\ 2013). 

Therefore, no biases in distance, evolutionary stage or missing fragments in \Nmm\ should be strongly affecting the results presented in this work.

\subsection{Trend of fragmentation level with density power-law index and density}

In this Section we study the relation between the core parameters inferred from the model and the fragmentation level, quantified through the number of millimeter sources within a field of view of 0.1~pc (in diameter), $N_\mathrm{mm}$, as in Palau \et\ (2013). 
We find no trend of $\beta$, $T_0$ and total mass of the core with the fragmentation level (Tables~\ref{tfit1} and \ref{tfit2}). 
Conversely, Fig.~\ref{fcorr} presents the relation between the density power-law index $p$, the central density (at 1000~AU) $\rho_0$, and the density within a radius of 0.05~pc, \nradius, with the fragmentation level. Fig.~\ref{fcorr}-a reveals a possible weak trend of \Nmm\ with the density power-law index, with low fragmentation levels found mainly in cores with steeper density profiles. Concerning \Nmm\ vs $\rho_0$, our sample splits up into two groups, cores with high fragmentation and high $\rho_0$, and cores with lower fragmentation and low $\rho_0$, suggesting that high central densities (in opposition to average density, as assessed in Palau \et\ 2013), could be related to the fragmentation level. Finally, there is a clear trend of high fragmentation levels in cores with high \nradius\ (Fig.~\ref{fcorr}-c). 
A closer inspection of the data using statistical tests, like the Spearman's $\rho$ and Kendall's $\tau$ rank correlation coefficients, indicates that $N_{\rm mm}$ and $p$ show a faint anti-correlation ($\rho \sim -0.49$, $\tau \sim -0.3$, Pearson's linear correlation coefficient $R \sim -0.3$). As for \nradius, it is strongly correlated with \Nmm\ ($\rho \sim 0.70$, $\tau \sim 0.54$), and the two parameters are well described by a linear relationship ($R\sim 0.90$).
These relations indicate that the high \nradius\ likely arise in cores with rather flat $p$ and also high $\rho_0$.

A trend of high fragmentation level for flat density profiles was already suggested theoretically by Myhill \& Kaula (1992). More recently, Girichidis \et\ (2011) show, by studying the evolution of collapsing non-magnetized turbulent massive cores for different initial density profiles, that cores with a flat density profile produce a high number of protostars, while concentrated density profiles rather yield one massive protostar in the core center. This can be understood in the framework of Jeans fragmentation, as the larger the density, the smaller the Jeans mass (for a given temperature) and the larger the number of expected fragments $N_\mathrm{Jeans}$. In Fig.~\ref{fcorr}-c we overplot the expected number of fragments estimated as $N_\mathrm{Jeans}=(M_\mathrm{0.05pc}\times CFE)/M_\mathrm{Jeans}$, with $M_\mathrm{Jeans}=0.6285\,[T/10\mathrm{K}]^{3/2}[n_\mathrm{H_2}/10^5\mathrm{cm}^{-3}]^{-1/2}$~\mo, for which we used $T\sim20$~K, and $M_\mathrm{0.05pc}$ being the mass within a radius of 0.05~pc (calculated directly from the density law inferred from our model). The Core Formation Efficiency (CFE) was adopted to be $17$\%, the average of the CFEs measured in Palau \et\ (2013). Thus, the trend of higher fragmentation for higher densities within a radius of 0.05~pc is consistent with what is expected for Jeans fragmentation. The relation found in this work between fragmentation level (which could be regarded as a proxy to cluster richness) and gas density is reminiscent of the relation found on larger spatial scales between cluster mass or protostellar density and gas surface density (\eg\ Gonz\'alez-L\'opezlira \et\ 2012;  Lombardi \et\ 2013; 	
Pflamm-Altenburg \et\ 2013) and further studies should be conducted in this direction.

\subsection{Physical processes affecting the density structure of Massive Dense Cores \label{sdaffectingdens}}

With the aim of shedding light on which physical conditions promote fragmentation in massive dense cores, in this Section we study if there is any relation between different physical processes and the density structure (density profile and central density) of massive dense cores.

\begin{figure}
\begin{center}
\begin{tabular}[b]{c}
    \epsfig{file=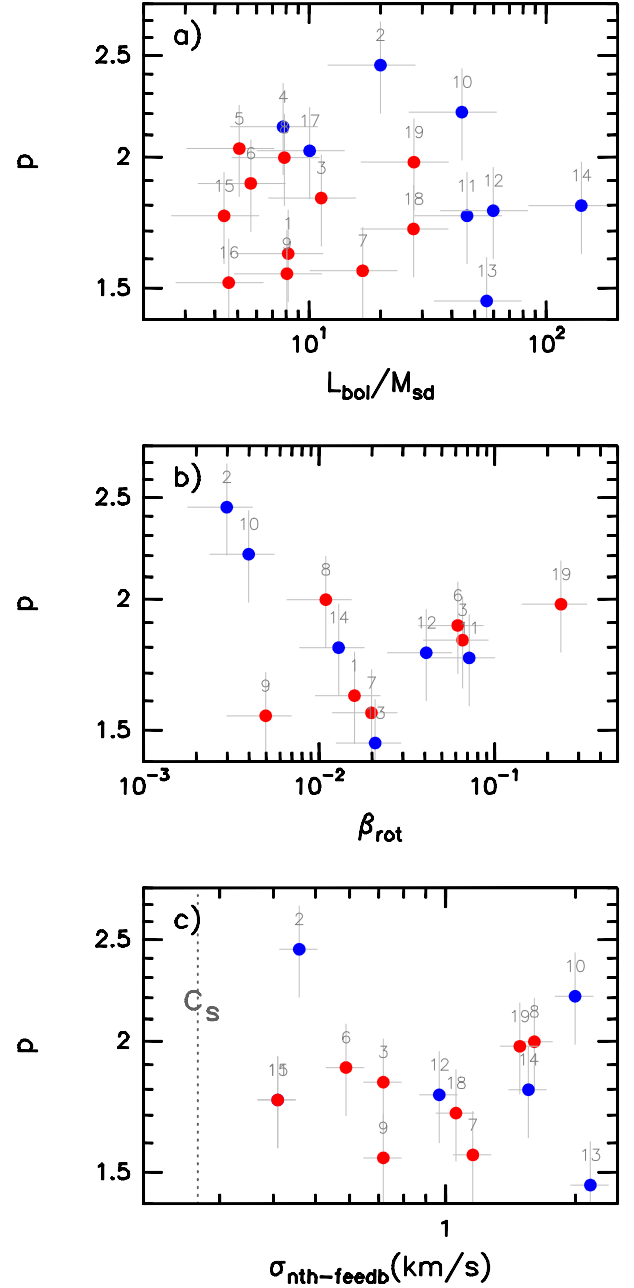, width=8cm, angle=0}\\
\end{tabular}
\caption{
a) Relation between the bolometric luminosity to the single-dish mass parameter $L_\mathrm{bol}/M_\mathrm{sd}$ and the density power law index.
b) Relation between rotational-to-gravitational energy $\beta_\mathrm{rot}$ and the density power law index.
c) Relation between non-thermal velocity dispersion (including outflow feedback) and density power-law index. Both plots include new cores after Palau \et\ (2013, see Section~\ref{sdaffectingdens}). 
In all panels, red dots correspond to cores with `confirmed' compressive turbulence or with hints of compressive turbulence, and blue dots correspond to cores with no assessment of the turbulence mode in the literature (see Table~\ref{tfit2} and Section~\ref{sdaffectingdens}). 
}
\label{fcorr-p}
\end{center}
\end{figure}

\paragraph{Time evolution} 
It has been shown both from observations and numerical simulations that structures within collapsing molecular clouds become more centrally peaked with time (\eg\ Smith \et\ 2009; Giannetti \et\ 2013). Therefore, the cores for which we measure higher values in the density power-law index $p$ could just correspond to the most evolved ones. However, we searched for any relation between $p$ and the ratio of the bolometric luminosity to the single-dish mass parameter (an evolutionary indicator used in Palau \et\ 2013\footnote{For DR21-OH we estimated a ratio of bolometric luminosity to single-dish mass of 27.8.}), and found no trend in our data (Fig.~\ref{fcorr-p}-a). 
This lack of correlation of $p$ with time probably results from the fact that our sample consists of cores at very similar evolutionary stages (Section~4.2), not properly sampling broad enough timescales, and suggests that other processes in addition to time evolution are probably also playing a role in determining the density structure of a dense core.

\begin{figure}
\begin{center}
\begin{tabular}[b]{c}
    \epsfig{file=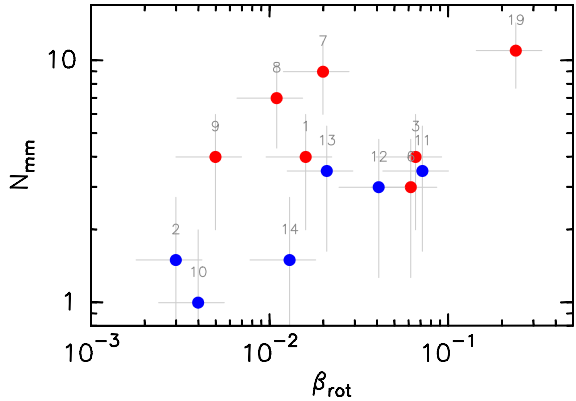, width=7.5cm, angle=0}\\
\end{tabular}
\caption{Relation between fragmentation level \Nmm\ and the rotational-to-gravitational energy ratio, after Palau \et\ (2013, i.e., including new measurements for OMC-1S and DR21-OH). Red dots correspond to cores with `confirmed' compressive turbulence or with hints of compressive turbulence, and blue dots correspond to cores with no assessment of the turbulence mode in the literature (see Table~\ref{tfit2} and Section~\ref{sdaffectingdens}). 
}
\label{fNmmbetarot}
\end{center}
\end{figure}

\paragraph{Rotational-to-gravitational energy ($\beta_\mathrm{rot}$)} 
The rotation of a massive dense core is expected to promote flattening of its density profile. In order to test this in our sample, we completed the values of rotational-to-gravitational energy ratios already estimated in Palau \et\ (2013) with the value for the core added in this work (DR21-OH), observed by Girart \et\ (2013) with similar conditions of sensitivity and spatial resolution as the sample of Palau \et\ (2013). Here we estimated \betar\ for DR21-OH following exactly the same method as in Palau \et\ (2013), \ie\ using \nh(1,1) data (Mangum \et\ 1992), and obtained a value of 0.24. We also estimated \betar\ for OMC-1S from the \nh(1,1) data by Wiseman \& Ho (1998). It is worth mentioning that these measurements of \betar\ must be regarded with caution as the velocity gradients observed might not be necessarily associated with rotation, but could arise from multiple velocity components or infall motions (\eg\ Lee \et\ 2013). Keeping this in mind, in Fig.~\ref{fcorr-p}-b we present the relation between \betar\ and $p$, showing a weak inverse trend. This suggests that the initial angular momentum alone might not be the only agent shaping the density profile of a core. We also present an updated version of the plot of \Nmm\ vs \betar\ (after Palau \et\ 2013) including these new measurements, and the updated plot, shown in Fig.~\ref{fNmmbetarot}, indicates a possible trend of \betar\ with fragmentation level, which deserves further studies.

\paragraph{Non-thermal velocity dispersion ($\sigma_\mathrm{non-th}$)} 
Palau \et\ (2013) estimate the non-thermal velocity dispersion from the linewidth measured in \nh(1,1) of a quiescent massive dense core in the immediate surroundings of the core where fragmentation was assessed. This can be regarded as a first approach to the initial turbulent dispersion. We completed the values of non-thermal `initial' velocity dispersions by adding two new measurements (CygX-N48 and DR21-OH by Mangum \et\ 1992, Table~\ref{tfit2}). However, the non-thermal velocity dispersions cover a small range of values, from 0.33 to 0.67~\kms, and did not show any relation with the fragmentation level.  We additionally study the relation of non-thermal velocity dispersion of the fragmenting cores (thus including feedback from outflows), covering a broader range from 0.41 to 2.16~\kms, with the density profile (Table~\ref{tfit2}, where we used the values from \nh(1,1) observed with the VLA: Torrelles \et\ 1989; Zhou \et\ 1990; Mangum \et\ 1992; Tieftrunk \et\ 1998; Wiseman \& Ho 1998; G\'omez \et\ 2003; S\'anchez-Monge \et\ 2013).
Since no clear trend is found (see Fig.~\ref{fcorr-p}-c and Table~\ref{tfit2}), this indicates that probably the density profile and the turbulent velocity dispersion are independent quantities. Finally, we estimated the virial masses and virial parameter ($M_\mathrm{vir}/M_\mathrm{0.05pc}$) using the non-thermal velocity dispersions of the fragmenting cores (\ie\ including feedback), and found no relation of these quantities with the density power-law index or the fragmentation level.

\paragraph{Turbulence mode: compressive vs solenoidal} 
In order to study if different turbulence modes could be affecting the density structure of massive dense cores, we searched the literature for any evidence of our massive dense cores being affected by expanding shells/bubbles created by nearby O-type stars, and our findings and references are listed in the two last columns of Table~\ref{tfit2}. Regions with hints in the literature of being associated with expanding shells or HII regions, such as bright-rimmed clouds, etc., or located at the meeting point of converging flows, are classified as `possible compressive', while regions for which the probability density function has been found to be consistent with compressive turbulence are classified as `confirmed compressive'. Both `possible' and `confirmed' compressive-turbulent cores (see Table~\ref{tfit2} to distinguish among both types) are assigned a `red' color in the figures, while the `blue' color corresponds to cores where the turbulence mode cannot be assessed with the current information (thus a solenoidal mode cannot be ruled out for `blue' cores). Fig.~\ref{fcorr} shows that `compressive-turbulent' cores tend to have large \Nmm, and cover a wide range of density power-law indices (from 1.5 to 2). In addition, cores with $\rho_0>10^{-16}$\,g\,cm$^{-3}$ and \nradius\ $>6\times10^5$\,cm$^{-3}$ fall all within the `compressive-turbulent' group (see Fig.~\ref{fcorr}-b,c). This could be consistent with compressive turbulence because this mode yields higher density contrast structures than the solenoidal mode (\eg\ Federrath \et\ 2008, 2010). Recently, Tremblin \et\ (2013) find that expanding HII regions or bubbles tend to induce steep density profiles in the cores at the bubble edges. Thus, the steep density profiles obtained in DR21-OH, AFGL\,5142, CB3 ($p\sim2$) could have been produced by the expanding ionization fronts from the massive stars in these regions (Hunter \et\ 1995; Schneider \et\ 2011; Sakai \et\ 2013).
A more detailed and uniform study of the turbulence mode in each core would be required to definitely assess the role of the different turbulence modes in setting the density structure of massive dense cores.

\begin{figure}
\begin{center}
\begin{tabular}[b]{c}
    \epsfig{file=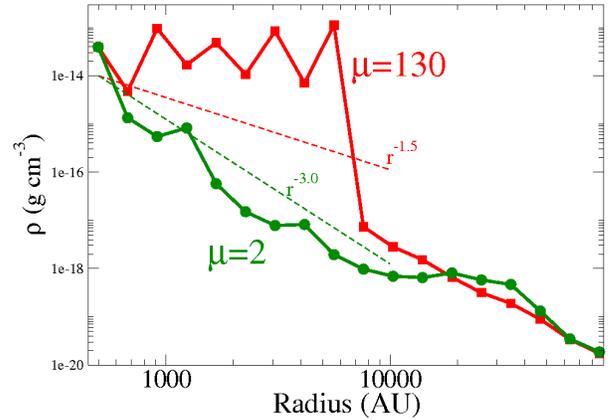 , width=5.5cm, angle=270}\\
\end{tabular}
\caption{Radial density profiles of $\mu=2$ (green line and circles) and $\mu=130$ (red line and squares) simulations of Commer\c con \et\ (2011). Note that for the magnetized case the core has a more concentrated profile than for the weakly magnetized case. The $r^{-3}$ and $r^{-1.5}$ profiles (dashed) are shown to guide the eye.}
\label{fsim}
\end{center}
\end{figure}

\paragraph{Magnetic field} 
Magnetic field could affect the density structure of a core because it helps removing angular momentum yielding rather steep density profiles. 
We tested this by using the magnetohydrodynamic simulations from Commer\c con \et\ (2011) of strongly ($\mu=2$, with $\mu$ being the mass-to-flux over the critical mass-to-flux ratio) and weakly ($\mu=130$) magnetized cores. These simulations were found to reproduce remarkably well our data (Palau \et\ 2013).
It is important to note that in these simulations the initial density profile is the same and rather flat for both $\mu=2$ and $\mu=130$ cases. We extracted the density profiles of Commer\c con \et\ (2011) simulations for $\mu=2$ and $\mu=130$ cases, by making an average of the density on concentric spherical shells weighted by the mass 
for each cell in the shell. The result is displayed in Fig.~\ref{fsim}, and shows that the $\mu=2$ case evolved into a more concentrated profile than the $\mu=130$ case, for radii up to $\sim0.05$~pc or 10000~AU.
This results from the fact that for the low magnetic field case ($\mu=130$), the fragmented area is large because of conservation of angular momentum, which gives a flatter density distribution at the center. And the opposite, strong magnetic braking (in the $\mu=2$ case) helps concentrate matter in the center and gives a steeper profile. 
In addition, previous theoretical work including ambipolar diffusion (\eg\ Lizano \& Shu 1989) also predict steep density profiles (density power-law indices of $\sim 2$).
Thus, the current theoretical and numerical works suggest a relation between the inferred density power-law index of massive dense cores and the magnetic field properties. 
Finally, we also used the simulations from Commer\c con \et\ (2011) to estimate the density within a radius of 0.05~pc for both $\mu=2$ and $\mu=130$, and find \nradius\ $\sim2.3\times10^5$~cm$^{-3}$ and $5.8\times10^5$~cm$^{-3}$, respectively. This, together with the number of millimeter sources that a typical interferometer would detect in each simulation (estimated in Palau \et\ 2013), allowed us to compare the results of these simulations to our \Nmm\ vs \nradius\ plot showing observational data (see Fig.~\ref{fcorr}-c). The values predicted by the simulations match remarkably well the trend seen in the observational data, suggesting that the magnetic field strength might be related to \nradius\ of massive dense cores, with small densities found in the strongly magnetized cores and large densities found in weakly magnetized cores. Overall, the agreement between Commer\c con \et\ (2011) simulations and our data is very good and suggests that magnetic field might be affecting the shape and density of massive dense cores, which in turn seems to be related to the fragmentation level.

\section{Conclusions}

In order to study the relation between the density profile of massive dense cores and the fragmentation level, we have fitted simultaneously the radial intensity profiles at 450 and 850~\mum\ and the SED of 19 massive dense cores for which their relative fragmentation level was known from previous works. Although the parameter used to estimate the fragmentation level is a lower limit, its relative value for the different cores is significant because it is assessed in a uniform way for all of them. The sample consists of massive dense cores undergoing active star formation and being at similar evolutionary stages. We modeled the cores as spherical envelopes with density and temperature decreasing with radius following power-laws, and found a weak trend of lower fragmentation for cores with steeper ($p\ga2.0$) density profiles. A correlation is found between the fragmentation level and the density within a radius of 0.05~pc (the radius where fragmentation was assessed), suggesting that high fragmentation levels might be related with flat cores and cores with high central densities. This correlation is consistent with Jeans fragmentation, as high densities yield lower Jeans masses.

While we find no clear relation between velocity gradients or velocity dispersion and the density power law index of our massive dense cores, cores with compressive turbulence seem to be associated with high central densities, thus promoting fragmentation. On the other hand, a comparison with radiation magnetohydrodynamical simulations suggests that a strong magnetic field could help shape steep density profiles in massive dense cores, for which low fragmentation levels are found. Thus, it is of crucial importance to study the relation of density profile, turbulence mode and magnetic field from both an observational an numerical point of view.

\acknowledgments
\begin{small}
The authors are grateful to the anonymous referee for a critical reading of the manuscript providing key comments that improved the paper.
A.P. is grateful to Tim Jenness, Luis Chavarria, and Floris van der Tak for prompt support regarding the JCMT submillimeter continuum data, to Ana Duarte-Cabral for sharing the Herschel photometry of Cygnus-X cores, to Christoph Federrath, Susana Lizano, Enrique V\'azquez-Semadeni, Javier Ballesteros-Paredes, and Andrew Myers for insightful discussions, and Carmen Ju\'arez and Jennifer Wiseman for sharing \nh\ VLA data.
A.P. and J.M.G. are supported by the Spanish MICINN grant AYA2011-30228-C03-02 (co-funded with FEDER funds), and by the AGAUR grant 2009SGR1172 (Catalonia). 
A.F. thanks the Spanish MINECO for funding support from grants CSD2009-00038 and AYA2012-32032.
G.B. is supported by an Italian Space Agency (ASI) fellowship under contract number I/005/11/0. 
L.A.Z. acknowledge the financial support from DGAPA, UNAM, and CONACyT, M\'exico.
This research used the facilities of the Canadian Astronomy Data Centre operated by the National Research Council of Canada with the support of the Canadian Space Agency.
\end{small}

\begin{appendix}



\begin{figure*}
\begin{center}
\begin{tabular}[b]{c}
    \epsfig{file=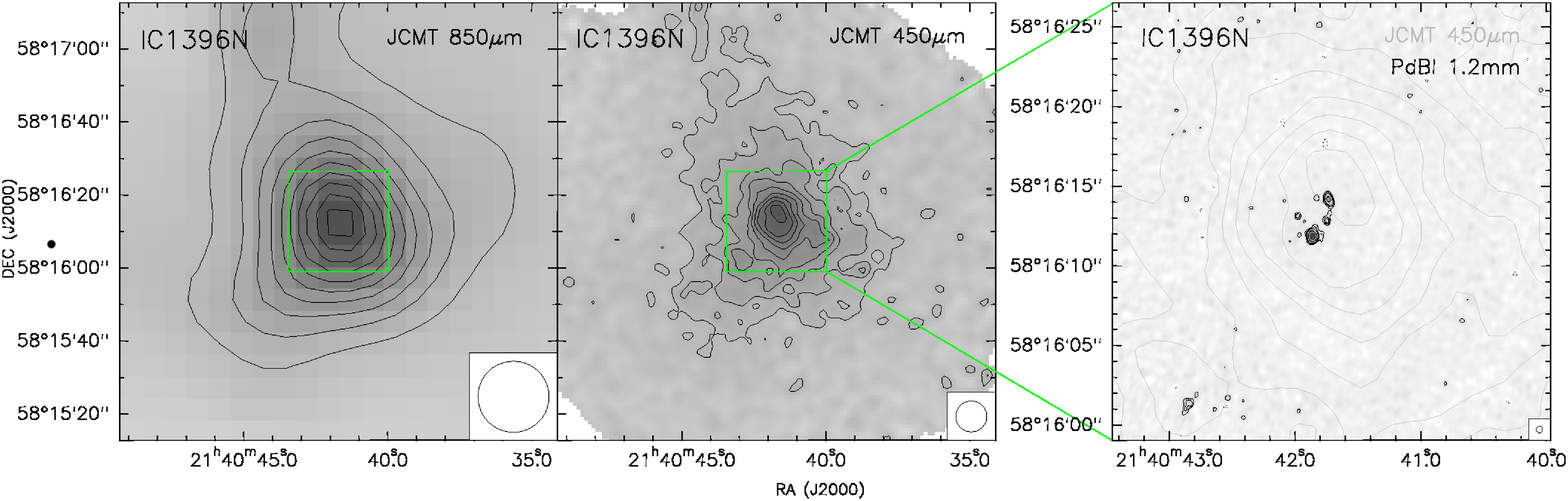, width=14.cm, angle=0}\\
    \epsfig{file=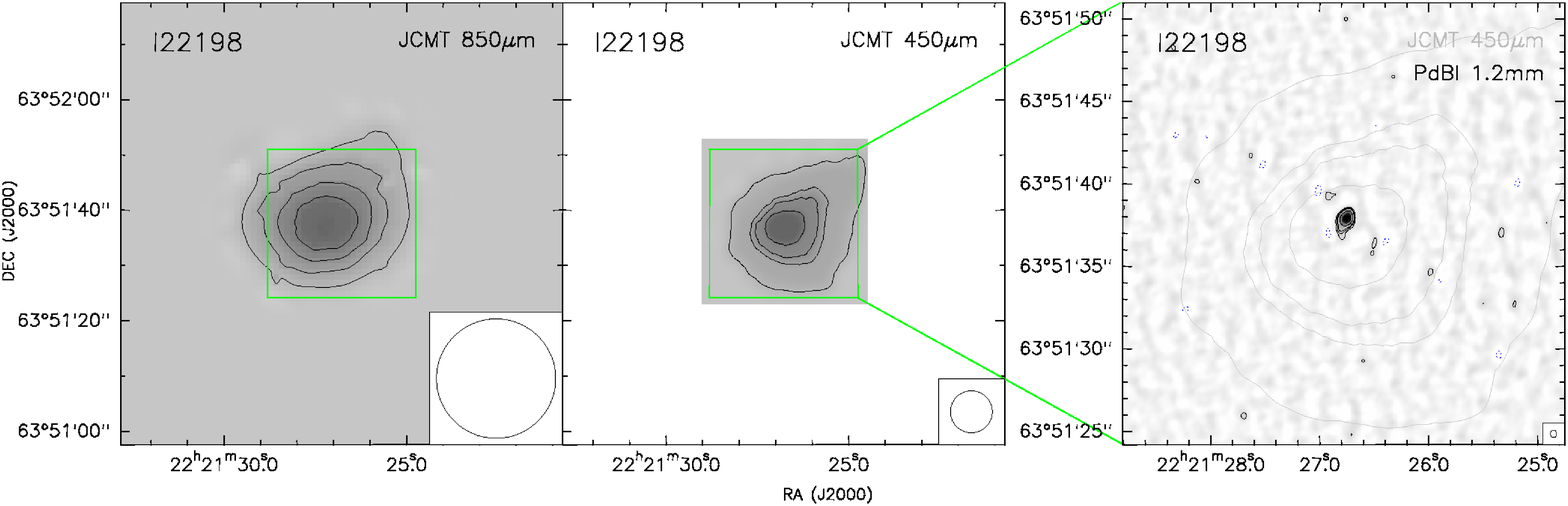, width=14.cm, angle=0}\\
    \epsfig{file=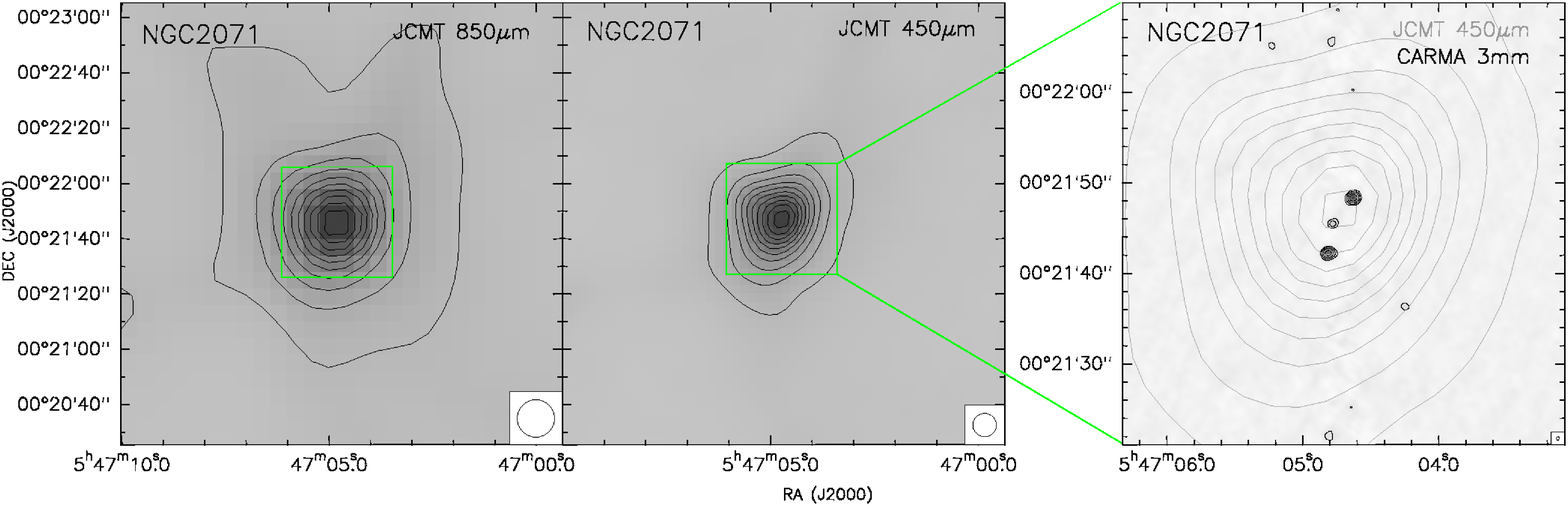, width=14.cm, angle=0}\\
    \epsfig{file=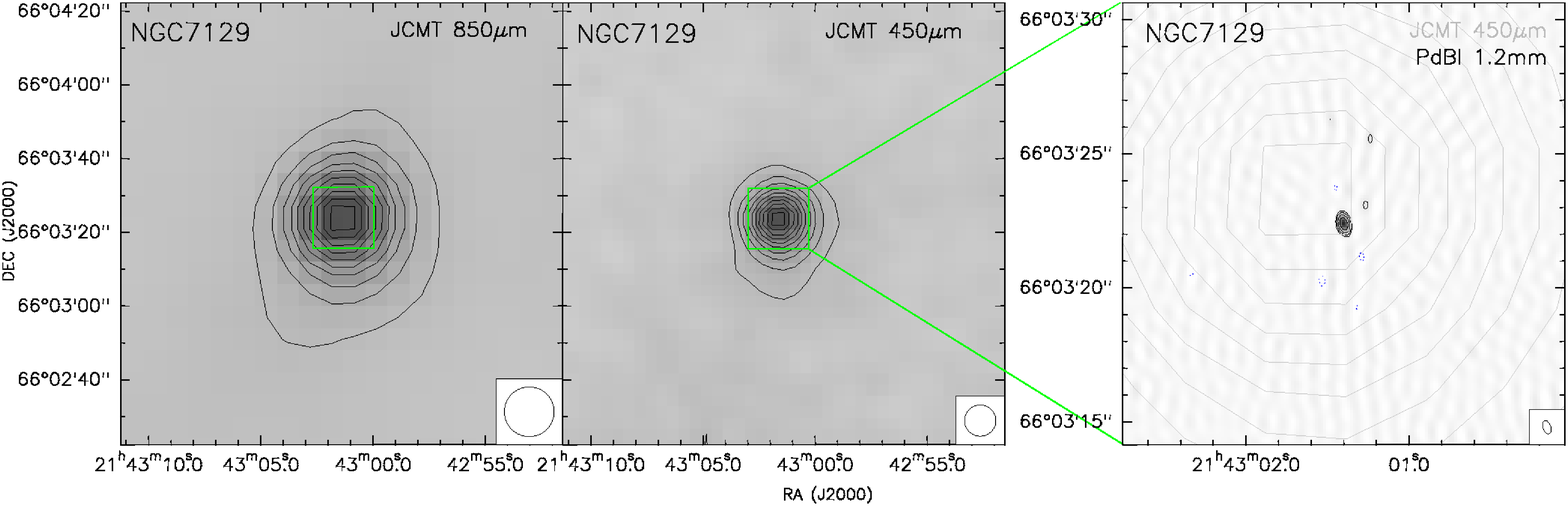, width=14.cm, angle=0}\\
    \epsfig{file=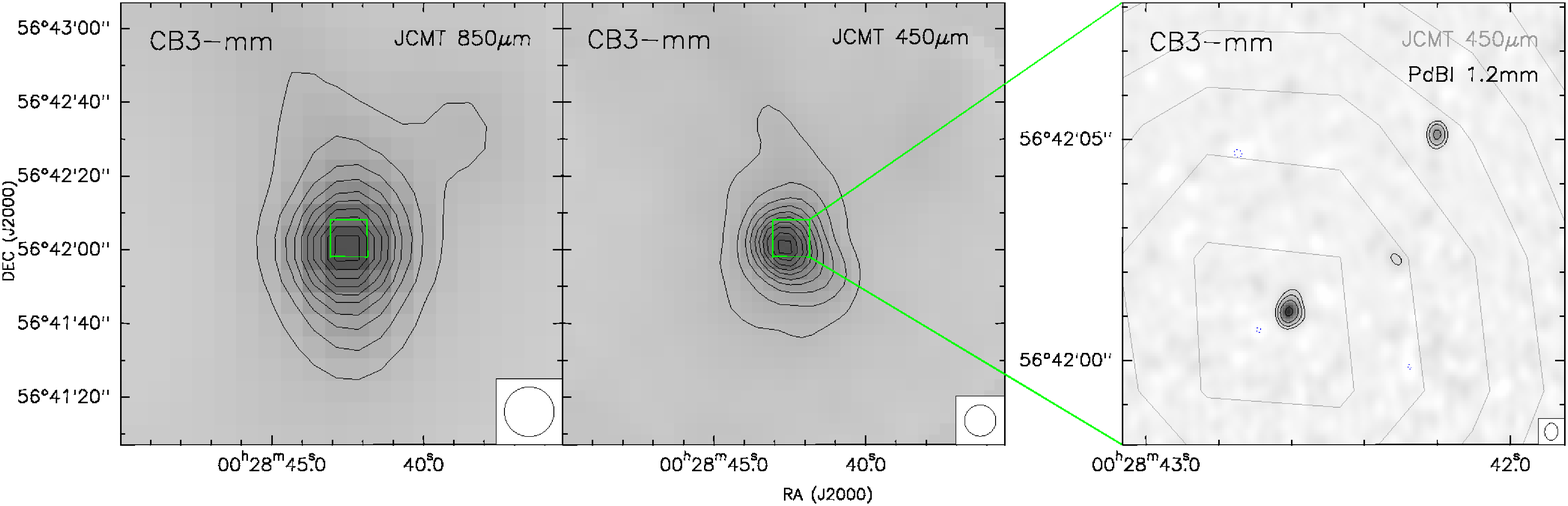, width=14.cm, angle=0}\\
\end{tabular}
\caption{Left and central panels: JCMT/SCUBA 850 and 450~\mum\ emission (Jenness \et\ 1995; Di Francesco \et\ 2008). For each source, the two green squares are $\sim0.1\times0.1$\,pc$^2$, the size of the mm interferometer map (right panel, Palau et al. 2013) revealing the fragmentation in each core. For IC1396N, NGC\,7129-FIRS2, and CB3-mm, contours of 850 (450)~\mum\ emission (left (middle) panels) are 8\% to 99\%, increasing in steps of 10\% of the peak value, 5.0 (25.3), 4.1 (48.9), and 2.1 (12.8)~\jpb, respectively. For I22198, contours of 850 (450)~\mum\ emission (left (middle) panel) are 20\% to 99\%, increasing in steps of 20\% of the peak value, 5.5 (9.8)~\jpb. For NGC\,2071, contours of 850 (450)~\mum\ emission (left (middle) panel) are 5\% to 99\%, increasing in steps of 10\% of the peak value, 13.1 (97.3)~\jpb.
Contours for interferometric images (right panels) are $-4$, 4, 8, 16, 32 and 64 times the rms noise of each map as given in Palau \et\ (2013).
}
\label{fobs1}
\end{center}
\end{figure*}

\begin{figure*}
\begin{center}
\begin{tabular}[b]{c}
    \epsfig{file=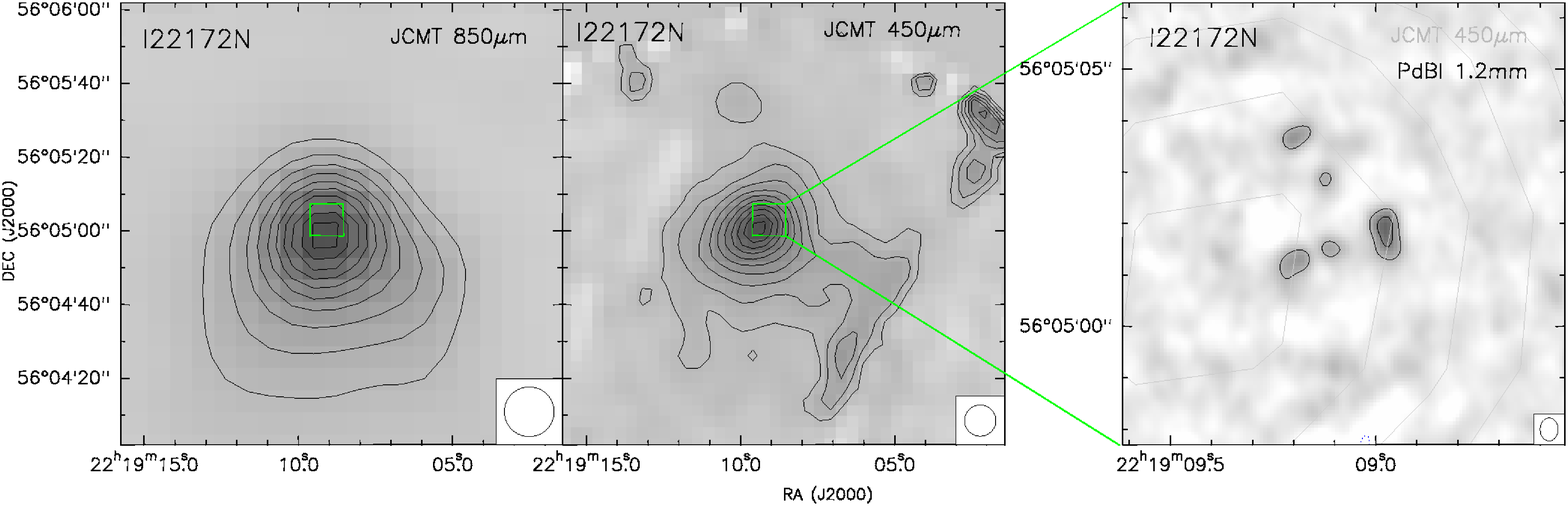, width=14.cm, angle=0}\\
    \epsfig{file=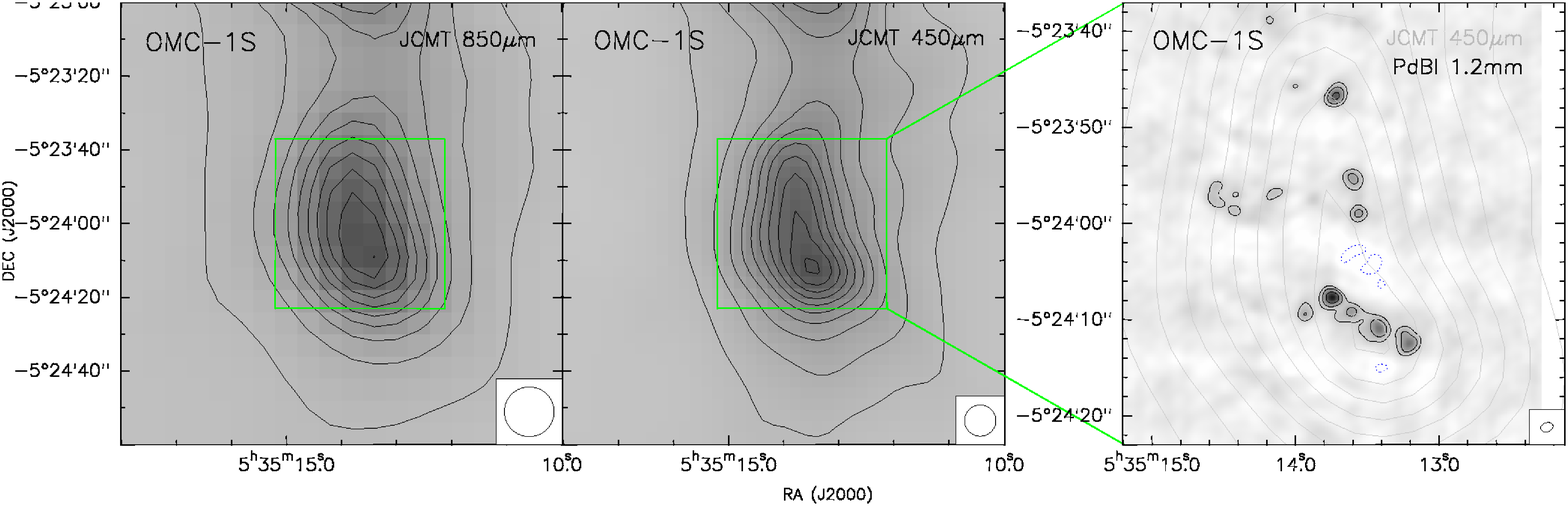, width=14.cm, angle=0}\\
    \epsfig{file=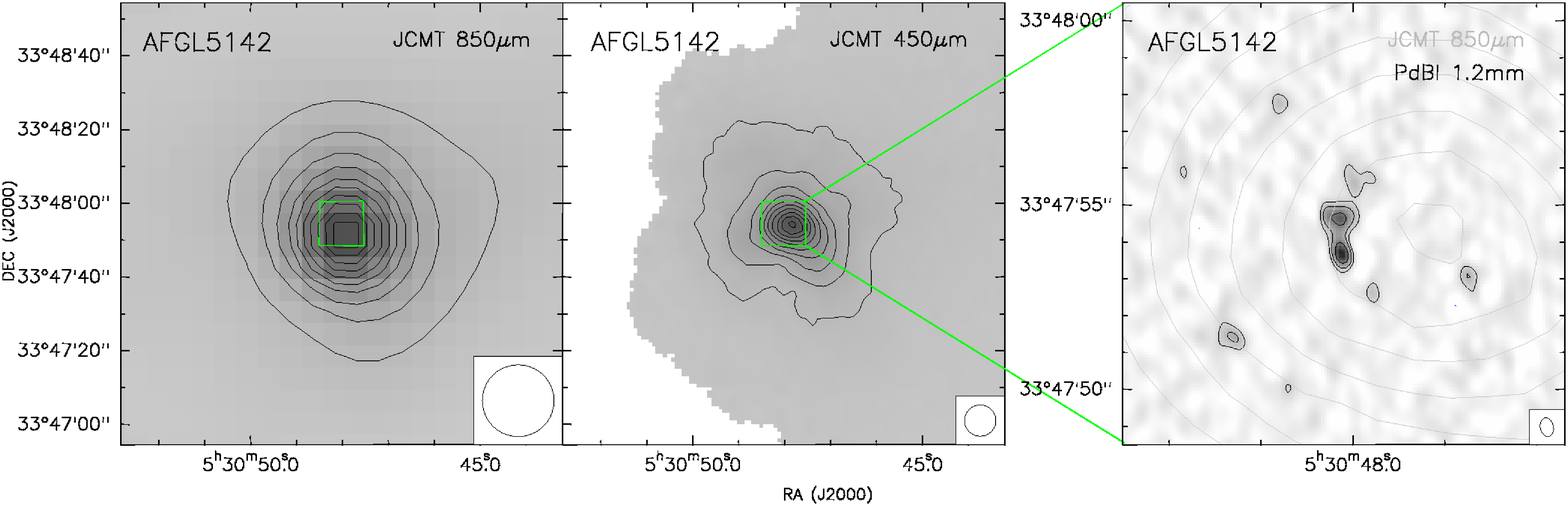, width=14.cm, angle=0}\\
    \epsfig{file=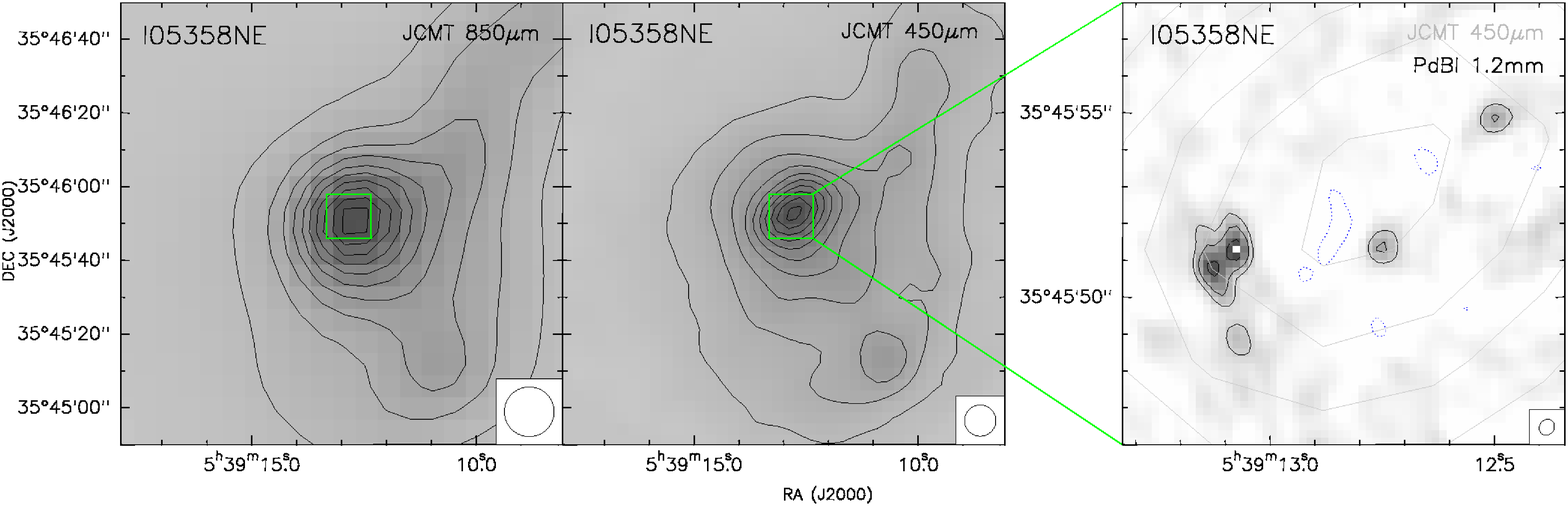, width=14.cm, angle=0}\\
    \epsfig{file=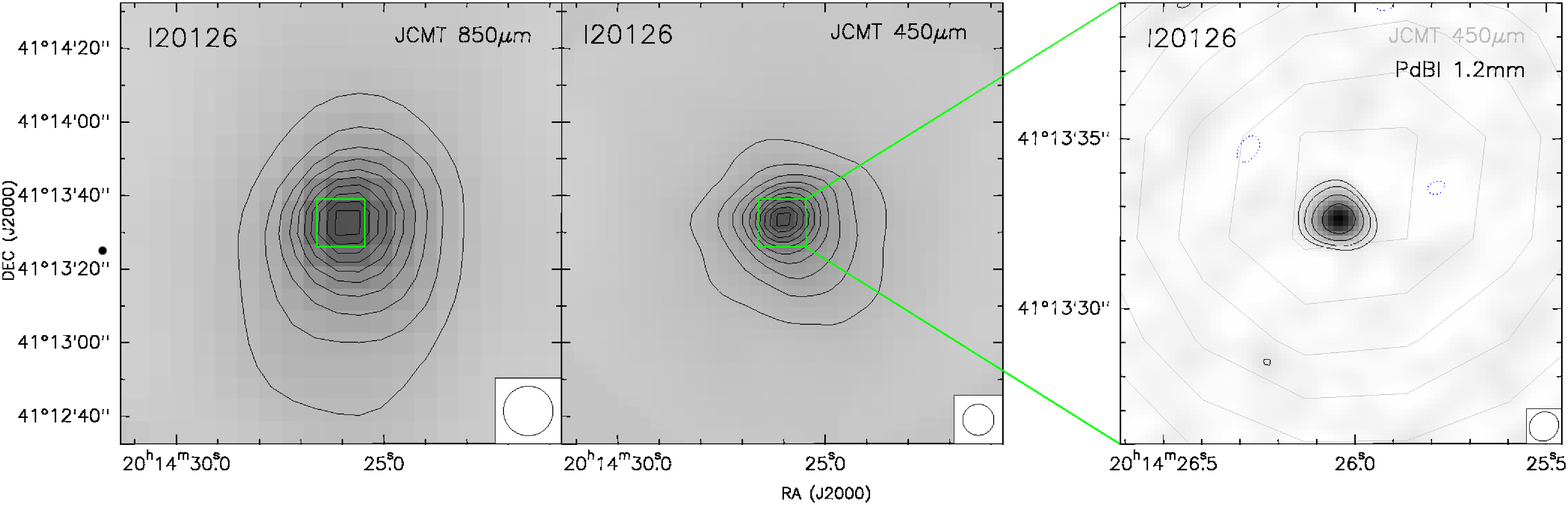, width=14.cm, angle=0}\\
\end{tabular}
\caption{Left and central panels: JCMT/SCUBA 850 and 450~\mum\ emission (Di Francesco \et\ 2008). For each source, the two green squares are $\sim0.1\times0.1$\,pc$^2$, the size of the mm interferometer map (right panel, Palau et al. 2013) revealing the fragmentation in each core.
For I22172N, OMC-1S, AFGL\,5142, I05358NE, and I20126, contours of 850 (450)~\mum\ emission (left (middle) panels) are 8\% to 99\%, increasing in steps of 10\% of the peak value, 2.0 (13.2), 73.5 (390), 9.5 (45.4), 7.9 (27.6), 6.1 (62.4)~\jpb, respectively. 
Contours for interferometric images (right panels) are $-4$, 4, 8, 16, 32 and 64 times the rms noise of each map as given in Palau \et\ (2013).
}
\label{fobs2}
\end{center}
\end{figure*}

\begin{figure*}
\begin{center}
\begin{tabular}[b]{c}
    \epsfig{file=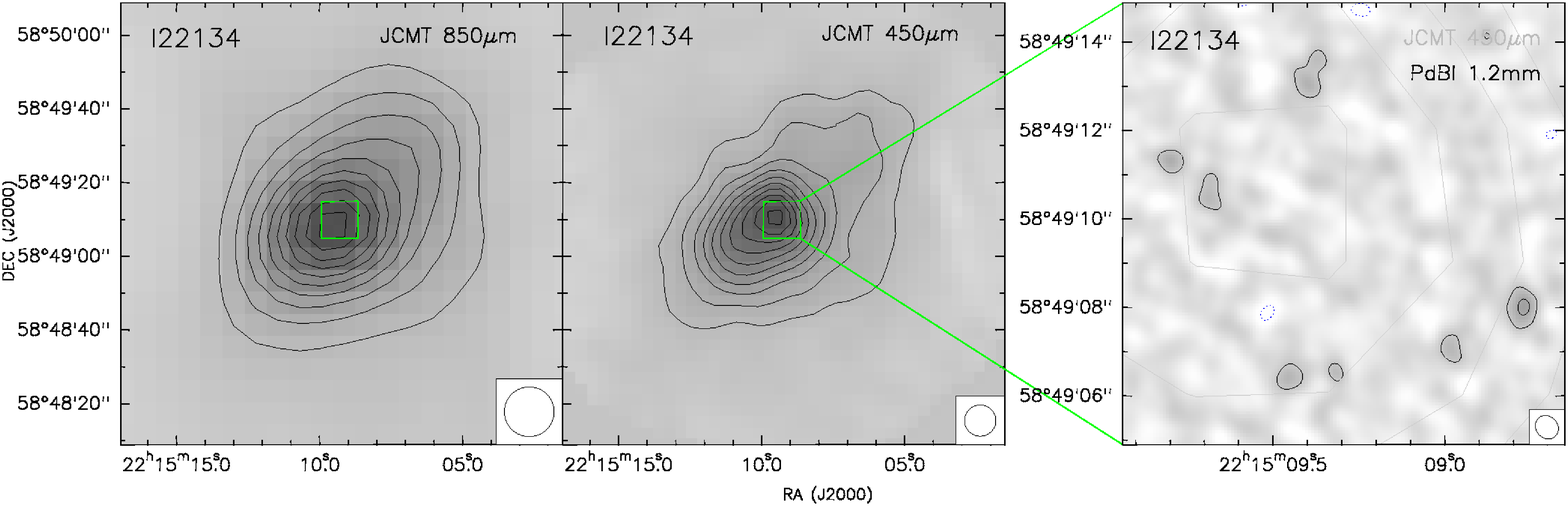, width=14cm, angle=0}\\
    \epsfig{file=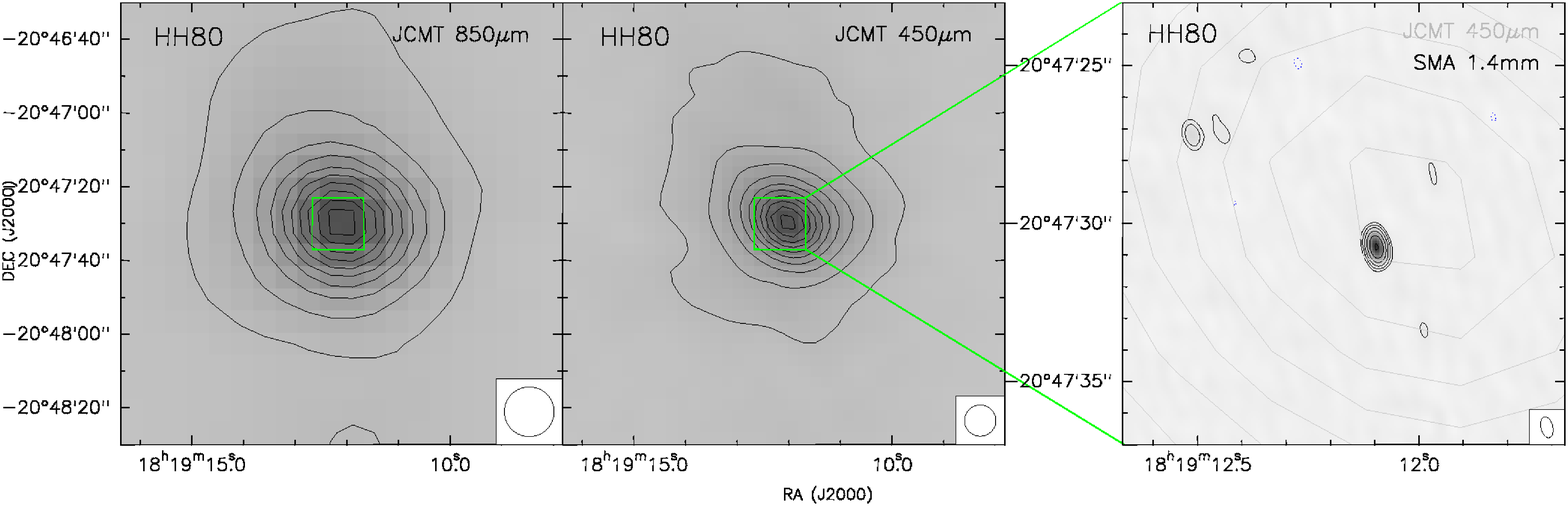, width=14cm, angle=0}\\
\end{tabular}
\caption{Left and central panels: JCMT/SCUBA 850 and 450~\mum\ emission (Di Francesco \et\ 2008). For each source, the two green squares are $\sim0.1\times0.1$\,pc$^2$, the size of the mm interferometer map (right panel, Palau et al. 2013) revealing the fragmentation in each core.
For I22134, HH80, W3IRS5, and AFGL\,2591, contours of 850 (450)~\mum\ emission (left (middle) panels) are 8\% to 99\%, increasing in steps of 10\% of the peak value,  2.6 (20.4), 8.3 (62.1), 18.0 (76.8), 7.9 (71.4)~\jpb, respectively. 
Contours for interferometric images (right panels) are $-4$, 4, 8, 16, 32, 64 and 128 times the rms noise of each map as given in Palau \et\ (2013).
NOTE: FIGURES FOR REGIONS W3IRS5, AFGL2591, CYGX-N53, CYGX-N12, CYGX-N63, CYGX-N48, AND DR21-OH, as well as Fig.~14 can be found in:  
http://www.ice.csic.es/files/palau/Palau14.pdf
}
\label{fobs3}
\end{center}
\end{figure*}

\end{appendix}


\end{document}